\documentclass[aps,prd,showpacs,nofootinbib,amssymb,preprint]{revtex4-1}
\pdfoutput=1
\usepackage{amssymb,graphicx,color,ulem,amsmath}
\usepackage{epsfig}

\begin{document}

\title{Polarization effects in the search for dark vector boson at $e^+e^-$ colliders }

\author{Fei-Fan Lee$^{a}$, Guey-Lin Lin$^{b}$, and Vo Quang Nhat $^{b}$}
\affiliation{$^{a}$ Department of Physics, Jimei University,
361021, Xiamen, Fujian province, P. R. China \\
$^{b}$Institute of Physics, National Chiao Tung University, Hsinchu 30010,
Taiwan
}
\date{\today}

\begin{abstract}
We argue that the search for dark vector boson through $e^+e^-\to Z_d\gamma$ can determine the Lorentz structure of $Z_dl^+l^-$ couplings with the detection of leptonic decays $Z_d\to l^+l^-$. We assume a general framework that the dark vector boson interacts with ordinary fermions through vector and axial-vector couplings. As a consequence of Ward-Takahashi identity, $Z_d$ is transversely polarized in the limit $m_{Z_d}\ll \sqrt{s}$. On the other hand,  the fraction of longitudinal $Z_d$ is non-negligible for $m_{Z_d}$ comparable to $\sqrt{s}$. Such polarization effects can be analyzed through angular distributions of final-state particles in $Z_d$ decays. 
Taking $l^{\pm}\equiv \mu^{\pm}$, we study the correlation between $Z_d$ angle relative to $e^-$ beam direction in $e^+e^-$ CM frame  and $\mu^-$ angle relative to the boost direction of $Z_d$ in $Z_d$ rest frame.
This correlation is shown to be useful for probing the Lorentz structure of $Z_dl^+l^-$ couplings.
We discuss the measurement of such correlation in Belle II detector, taking into account the detector acceptance and energy resolution. 
\end{abstract}

\maketitle
 
 \newpage
 \section{Introduction}
 Searching  for dark matter (DM) is one of the major endeavors in the present day particle physics community. The efforts in direct and indirect detections as well as productions of DMs in LHC so far have not produced positive results. Recently 
 there are growing interests to search for DM related phenomena with huge statistics and high precision measurements. These phenomena involve the hidden sector~\cite{Holdom:1985ag,Galison:1983pa,Foot:2004pa,Feldman:2006wd,ArkaniHamed:2008qn,Pospelov:2008jd}, which is assumed to 
 interact with Standard Model (SM) particles through certain messengers. A popular proposal for such a messenger particle is the so called dark photon, which mixes with the 
 $U(1)$ hypercharge field $B_{\mu}$ in SM,
 \begin{eqnarray}
 \mathcal{L}_{\rm{gauge}}=-\dfrac{1}{4}B_{\mu\nu}B^{\mu\nu}+\dfrac{1}{2}\dfrac{\varepsilon_{\gamma}}{\rm{cos}\theta_{W}}B_{\mu\nu}A^{\prime}_{\mu\nu}-\frac{1}{4}A^{\prime}_{\mu\nu}A^{\prime\mu\nu},
 \label{eq:Lag_gauge}
 \end{eqnarray}
 where $A^{\prime}_{\mu}$ is the dark photon field, and $A^{\prime}_{\mu\nu}\equiv \partial_{\mu}A^{\prime}_{\nu}-\partial_{\nu}A^{\prime}_{\mu}$. 
 The above mixing induces electromagnetic couplings, $\mathcal{L}_{\rm em}=\varepsilon_{\gamma}eJ^{\mu}_{\rm em}A^{\prime}_{\mu}$, between the dark photon and SM fermions, which generate rich 
 phenomenology~\cite{Alexander:2016aln}. 
 On the other hand, the neutral current couplings between the same set of particles are further suppressed by the factor $m_{A^{\prime}}^2/m_Z^2$ for $m_{A^{\prime}}\ll m_Z$ with $m_{A^{\prime}}$ the dark photon mass. 
 However, independent neutral current couplings can be generated through mass mixing between the messenger particle and the $Z$ boson~\cite{Babu:1997st,Davoudiasl:2012ag,Davoudiasl:2013aya}.
 In this case, the messenger particle is often referred to as $Z^{\prime}$ boson. 
 The mass mixing term $\delta m^2 Z^{\prime}_{\mu}Z^{\mu}$ can induce
 neutral current couplings $\mathcal{L}_{\rm NC}=(g\varepsilon_Z /\cos\theta_{\rm W}) J^{\mu}_{\rm NC}Z^{\prime}_{\mu}$ with $\varepsilon_Z\equiv \delta m^2/m^2_{Z}$. 
For a general scenario that both kinetic and mass mixings are present, the interactions between dark boson and SM fermions are given by 
  \begin{equation}
 {\cal L}_{\rm int}=\left(\varepsilon_{\gamma}eJ^{\mu}_{\rm em}+\varepsilon_Z \frac{g}{\cos\theta_{\rm W}}J^{\mu}_{\rm NC}\right)Z_{d,\mu},
 \label{eq:INT}
 \end{equation}	 
 with $Z_d$ the dark boson, which is the generalization of $A^{\prime}$ and $Z^{\prime}$\footnote{Strictly speaking, $\varepsilon_Z$ in Eq.~(\ref{eq:INT}) is $\delta m^2/m^2_{Z^0}$ with $m_{Z^0}$ the $Z$ boson mass before mass matrix diagonalization. Nevertheless, for numerical calculations, we take $\varepsilon_Z\equiv \delta m^2/m^2_{Z}$ as the correction is of higher orders.}. 
 
 We note that the physical masses of $Z_d$ and $Z$ are given by diagonalizing of the following mass matrix, 
 \begin{equation}
 \mathbf{M_0^2}=\begin{pmatrix}
m^2_{Z^0} & -\varepsilon_Z m_{Z^0}^2 \\
-\varepsilon_Z m_{Z^0}^2 & m^2_{Z_d^0} 
\end{pmatrix}
 +\begin{pmatrix}
0 & -\varepsilon_\gamma \tan\theta_{\rm W} m^2_{Z^0} \\
-\varepsilon_\gamma \tan\theta_{\rm W} m^2_{Z^0} &  (2\varepsilon_\gamma\varepsilon_Z \tan\theta_{\rm W} m^2_{Z^0}+\varepsilon^2_\gamma\tan\theta_{\rm W}^2 m^2_{Z^0})
\end{pmatrix},
 \end{equation} 
where the second term arises from the diagonalization of kinetic Lagrangian $ \mathcal{L}_{\rm{gauge}}$.
To ensure the positivity of physical masses, one requires the determinant of $\mathbf{M_0^2}$ be positive, i.e., ${\rm det} (\mathbf{M_0^2})= (m^2_{Z_d^0}m^2_{Z^0}-(\delta m^2)^2)>0$. We note that the second term of the mass matrix does not give additional contribution to the determinant.  
This condition constrains models that generate masses for $Z$, $Z_d$, and the mass mixing between the two. For our interested scenario $\varepsilon_Z\simeq \varepsilon_\gamma$ with the latter in the range $(10^{-3}-10^{-4})$, i.e., 
current limits from dark photon searches at $e^+e^-$ colliders, $m_{Z^0_d}$ 
should be greater than $(10-100)$ MeV. In Ref.~\cite{Davoudiasl:2012ag},  two Higgs doublets and one scalar Higgs singlet are introduced such that $m^2_{Z^0}=g^2(v_1^2+v_2^2)/(4\cos^2\theta_{\rm W})$, 
$m^2_{Z_d^0}=g_d^2(v_2^2+v_d^2)$, and $\delta m^2=g_dgv_2^2/(2\cos\theta_{\rm W})$. Here $g_d$ is the dark boson gauge coupling to the second Higgs doublet and the scalar Higgs singlet, while $v_1 (v_2)$, and $v_d$  are vacuum expectation 
values of neutral scalar in the first (second) Higgs doublet and that of the scalar Higgs singlet.  The above-mentioned lower limit on $m_{Z_d}$ requires $g_d> 4\times (10^{-5}-10^{-4})$ if $(v_d^2+v_2^2)\simeq (v_1^2+v_2^2)\simeq (246 \ {\rm GeV})^2$. 

The search for the light vector boson with the reaction $e^+e^-\to Z_d\gamma$ has been proposed before~\cite{Boehm:2003hm,Borodatchenkova:2005ct,Fayet:2007ua}. Particularly, there exist phenomenological studies on dark sectors  under the environment of $e^+e^-$ colliders~\cite{Batell:2009yf,Essig:2009nc,Reece:2009un,Essig:2013vha,Karliner:2015tga,Araki:2017wyg,He:2017ord,He:2017zzr,Jiang:2018jqp}. 
Along this line, the experimental searches for $Z_d$ proceed through the detection of visible and invisible $Z_d$ decays. 
The visible mode requires a full reconstruction of $Z_d$ peak through measuring the energy and momentum of lepton or light hadron pairs from $Z_d$ decays~
\cite{Babusci:2014sta,Lees:2014xha,Anastasi:2015qla,Anastasi:2016ktq,Ablikim:2017aab,Anastasi:2018azp}, while
 the invisible mode looks for the peak of missing mass at $m_{Z_d}$ given by $M_{\rm mass}^2=(P_{e^-}+P_{e^{+}}-P_{\gamma})^2$~\cite{Lees:2017lec}. We note that both phenomenological and experimental studies mentioned above consider only the dark photon scenario, i.e., $Z_d$ interacts with SM fermions only via electromagnetic current. On the other hand, since neutral-current coupling is also possible, it is of great importance to simultaneously detect $Z_d$ and measure the Lorentz structure of its coupling to SM fermions. To determine the relative strengths of vector and axial-vector couplings, such as the ratio $g_{f,A}/g_{f,V}$ in the generic structure $e\varepsilon \bar{f}(g_{f,V}\gamma_{\mu}+g_{f,A}\gamma_{\mu}\gamma_5)fZ_d^{\mu}$, it is necessary to measure the angular distributions of final-state fermions from $Z_d$ decays\footnote{Here we choose the normalization $g_{f,V}^2+g_{f,A}^2=1$.}. 

The dark vector boson $Z_d$ produced by $e^+ e^-\to Z_d \gamma$ is polarized. In fact, $Z_d$ must be in one of the transversely polarized states in the limit $\sqrt{s}\gg m_{Z_d}$. This is a direct consequence of Ward-Takahashi identity~\cite{WT} to be elaborated in the next session. Furthermore, with the presence of both  $g_{f,A}$ and $g_{f,V}$, parity symmetry is broken. 
Hence there exists a forward-backward asymmetry for the production of each transversely polarized $Z_d$ state, while the production of longitudinal $Z_d$ is forward-backward symmetric. 
The magnitude of the above asymmetry is directly related to the degree of parity violation, characterized by the parameter $\rho\equiv 4g_{f,A}g_{f,V}$ under the normalization $g_{f,V}^2+g_{f,A}^2=1$.
For a fixed $\rho$, the asymmetry reaches to the maximum for $m_{Z_d}/\sqrt{s}\to 0$. 
Besides the asymmetry in the production of  transversely polarized $Z_d$ state, there is also forward-backward asymmetry for the angular distributions of final-state fermions from $Z_d$ decays, which is also controlled 
by the same parameter $\rho$.   
Hence the correlation between these two asymmetries can be exploited to probe $\rho$.
           
The most sensitive search for $Z_d$ through the visible mode $e^+e^-\to Z_d\gamma \to e^+e^-\gamma, \ \mu^+\mu^-\gamma$ is performed by BaBar~\cite{Lees:2014xha}. Using $514$ fb$^{-1}$ of data, the upper limits on the mixing parameter $\varepsilon$ is $10^{-4}-10^{-3}$ for $m_{Z_d}$ between $0.02$ GeV and $10.2$ GeV. Comparable sensitivity to $\varepsilon$ is expected at Belle II~\cite{Kou:2018nap,Abe:2010gxa,Brodzicka:2012jm,Bevan:2014iga} with $500$ fb$^{-1}$ of integrated luminosity.     
Belle II is an electron-positron collider experiment running at the SuperKEKB accelerator. It is a next-generation B-factory experiment aiming to record a dataset of 50 $\rm{ab^{-1}}$. In this article we focus on the prospect of detecting $Z_d$ and measuring the parity violation parameter in its interaction with SM fermions with 
$e^+e^- \to Z_d\gamma$ followed by $Z_d\to \mu^+\mu^-$ decay at Belle II. Backgrounds to this process are QED process $e^+e^-\to \mu^+\mu^-\gamma$~\footnote{Here we neglect the $Z$ boson exchange diagrams
since their entire contributions to the total cross section is less than $1\%$ from our numerical studies. }  and the resonant production process $e^+e^-\to \gamma X$ [$X=J/\psi, \ \psi(2S), \ \Upsilon(1S), \ \Upsilon(2S)$] followed by $X\to \mu^+\mu^-$. We will not consider the decay mode $Z_d\to e^+e^-$ in this article since backgrounds to this mode are more complicated, including $e^+e^-\to e^+e^-(\gamma)$ and $e^+e^-\to \gamma\gamma(\gamma)$. Since we are mainly interested in probing the parity violation effect by $Z_df\bar{f}$ coupling, the study on $Z_d\to \mu^+\mu^-$ is sufficient to make our point.  
We note that there are recent interests in the signals for $17$ MeV protophobic vector boson~\cite{Feng:2016jff,Feng:2016ysn,Feng:2020mbt} motivated from anomalies 
in $^8$Be and $^4$He nuclear transitions~\cite{Krasznahorkay:2015iga,Krasznahorkay:2019lyl}. Searching for vector boson in $e^+e^-$ colliders for this particular parameter range has been proposed~\cite{Jiang:2018uhs,Alikyhanov:2017cp}. Although we shall not focus on such a specific scenario, we do notice that the protophobic vector boson interacts with the electron through both vector and axial-vector couplings. However parity violation effects resulting from the 
presence of both couplings were not considered in the above analyses. For an earlier candidate of light neutral gauge boson~\cite{Fayet:1977yc}, its parity violating effects to low energy neutral current processes were studied quite some time ago~\cite{Fayet:1980ss}. Among those low energy processes, the search for atomic parity violations~\cite{Bouchiat:1974kt,Bouchiat:1979cq} still attracts high attentions in recent years. For phenomenological discussions on this issue under the current dark boson scenario, see Refs.~\cite{Davoudiasl:2012ag,Abdullah:2018ykz}.              
 
This article is organized as follows. In Section II, we present the polarized differential cross section of $e^+e^- \to Z_d\gamma$ for different $Z_d$ polarizations. For $\sqrt{s}\gg m_{Z_d}$, 
we show that the production of longitudinal $Z_d$ is suppressed due to Ward-Takahashi identity, i.e., $Z_d$ is transversely polarized in such a limit. In Section III, we discuss the method for probing the parity violation parameter $\rho$ in $e^+e^-$ colliders. We present angular distributions of leptons arising from polarized $Z_d$ decays. Combining with angular distributions of $Z_d$ in production process, we construct the double angular distribution for the signal process $e^+e^-\to \gamma Z_d\to \gamma l^+ l^-$. It will be shown that this double angular distribution depends on $\rho^2$ rather than $\rho$. We bin the signal events according to the sign of $J\equiv \cos\theta\times \cos\theta_d$ where $\theta$ is the angle of $Z_d$ with respect to the $e^-$ direction 
in $e^+e^-$ CM frame while $\theta_d$ is the helicity angle of lepton arising from $Z_d$ decay. The asymmetry ${\mathcal A}_{\rm PN}\equiv (S(J>0)-S(J<0))/(S(J>0)+S(J<0))$ with $S$ the number of signal events will be shown to be proportional to $\rho^2$, so that it directly reflects the degree of parity violation.
In Section IV, event numbers of $e^+e^-\to \gamma\mu^+\mu^-$ from signal and background are calculated with specific integrated luminosity in Belle II detector, taking into account the detector acceptance and energy resolutions. 
We also calculate the asymmetry parameter ${\mathcal A}_{\rm PN}$ which depends on the detector acceptance. It will be shown that the simultaneous fitting to $J>0$ and $J<0$ event bins should improve the significance 
of dark boson detection from simply counting the total event excess. The degree of improvement is closely related to  ${\mathcal A}_{\rm PN}$. In addition, 
it is possible to measure ${\mathcal A}_{\rm PN}$ in Belle II detector. We estimate the statistical errors for such measurements under Belle II design integrated luminosity. 
We summarize and conclude in Section V.  
 
\section{The polarized dark boson production cross section }
\subsection{Ward-Takahashi identity and the polarization of $Z_d$ }
Let us write the amplitude for $e^-(p_1)+e^+(p_2)\to Z_d(k_1)+\gamma (k_2)$ as
${\cal M}\equiv {\cal M}_{\mu}\epsilon^{\mu}(k_1)$ with $\epsilon^{\mu}(k_1)$ the polarization vector of $Z_d$. Here $\cal{M}_{\mu}$ contains the photon polarization vector. In the case that $Z_d$ is longitudinal, one has $\epsilon^{\mu}(k_1)=(|\vec{k}_1|,E_{Z_d}\hat{k}_1)/m_{Z_d}$. In the limit that $\sqrt{s}\gg m_{Z_d}$, i.e., $Z_d$ is ultra relativistic, one has $\epsilon_{Z_d}^{\mu}=k_1^{\mu}/m_{Z_d}+{\cal O} (m_{Z_d}/E_{Z_d})$. Hence ${\cal M}={\cal M}_{\mu}k_1^{\mu}/m_{Z_d} + {\cal O}(m_{Z_d}/E_{Z_d})$. However, ${\cal M}_{\mu}k_1^{\mu}=0$ in the limit $m_e\to 0$ as implied by Ward-Takahashi identity~\cite{WT}. Therefore the amplitude for a longitudinal polarized $Z_d$ is of the order $m_{Z_d}/\sqrt{s}$.

\subsection{Explicit demonstration of ${\cal M}_{\vert\vert}$ suppression}
The square of $e^-(p_1)+e^+(p_2)\to Z_d(k_1)+\gamma (k_2)$ amplitude for a given $Z_d$ polarization can be expressed as follows:   
\begin{eqnarray}
\frac{\vert \bar{{\cal M}}\vert ^2}{4}&=&16\pi^2\alpha^2\varepsilon^2(g_{f,V}^2+g_{f,A}^2)\bigg[\frac{u}{t}+\frac{t}{u}+\frac{2m_{Z_d}^2}{tu}\Big(s-2(p_1\cdot \epsilon^*)(p_1\cdot \epsilon)
-2(p_2\cdot \epsilon^*)(p_2\cdot \epsilon)\Big)\bigg] \nonumber \\
&-&64i\pi^2\alpha^2 \varepsilon^2 g_{f,V}\cdot g_{f,A} \times \Big( \frac{1}{t}\Big)\times \epsilon^{\rho\beta\sigma\nu}p_{1,\rho}(p_{2,\sigma}-k_{1,\sigma}) \epsilon_{\beta}\epsilon^*_{\nu}\nonumber \\
&+&64i\pi^2\alpha^2 \varepsilon^2 g_{f,V}\cdot g_{f,A} \times \Big( \frac{1}{u}\Big)\times \epsilon^{\rho\beta\sigma\nu}p_{2,\rho}(p_{1,\sigma}-k_{1,\sigma}) \epsilon_{\beta}\epsilon^*_{\nu}\nonumber \\
&-&128i\pi^2\alpha^2 \varepsilon^2 g_{f,V}\cdot g_{f,A}\times \Big( \frac{1}{tu}\Big) \times p_{2,\sigma}p_{1,\lambda}k_{1,\rho}\times \epsilon^{\rho\sigma\lambda\nu}\Big((p_2\cdot \epsilon)
\epsilon^*_{\nu}+(p_2\cdot \epsilon^*)\epsilon_{\nu}\Big) \nonumber \\
&+&128i\pi^2\alpha^2 \varepsilon^2 g_{f,V}\cdot g_{f,A}\times \Big( \frac{1}{tu}\Big) \times p_{2,\sigma}p_{1,\lambda}k_{1,\rho}\times \epsilon^{\rho\sigma\lambda\nu}\Big((p_1\cdot \epsilon)
\epsilon^*_{\nu}+(p_1\cdot \epsilon^*)\epsilon_{\nu}\Big), 
\label{eq:amplitude}
\end{eqnarray}
where $\bar{{\cal M}}$ is the amplitude with the polarizations of initial fermions and final-state photon summed, $\alpha$ is the fine-structure constant, $m_{Z_d}$ and $\epsilon_{\mu}$ are dark boson mass and polarization vector, respectively, $s$, $t$, and $u$ are Mandelstam variables. It is clear that those terms proportional to 
$g_{f,V}\cdot g_{f,A}$ vanish by summing the $Z_d$ polarization, $\sum_{\lambda}{\epsilon_{\mu}^{\lambda}\epsilon_{\nu}^{*\lambda}}=-g_{\mu\nu}+k_{1,\mu}k_{1,\nu}/m_{Z_d}^2$. 
In the center of momentum (CM) frame of colliding electrons and positrons, the momenta of initial and final-state particles are given by 
\begin{eqnarray}
p_1^{\mu}&=& (E, \ 0, \ 0, \ +E), \nonumber \\
p_2^{\mu}&=& (E, \ 0, \ 0, \ -E), \nonumber \\
k_1^{\mu} &=& (E_{Z_d}, \ \omega \sin\theta, \ 0, \ \omega \cos\theta), \nonumber \\
k_2^{\mu} &=& (\omega, \ -\omega\sin\theta, \ 0, \ -\omega \cos\theta),
\end{eqnarray} 
\begin{figure}[htbp]
	\begin{center}
		\includegraphics[width=12cm]{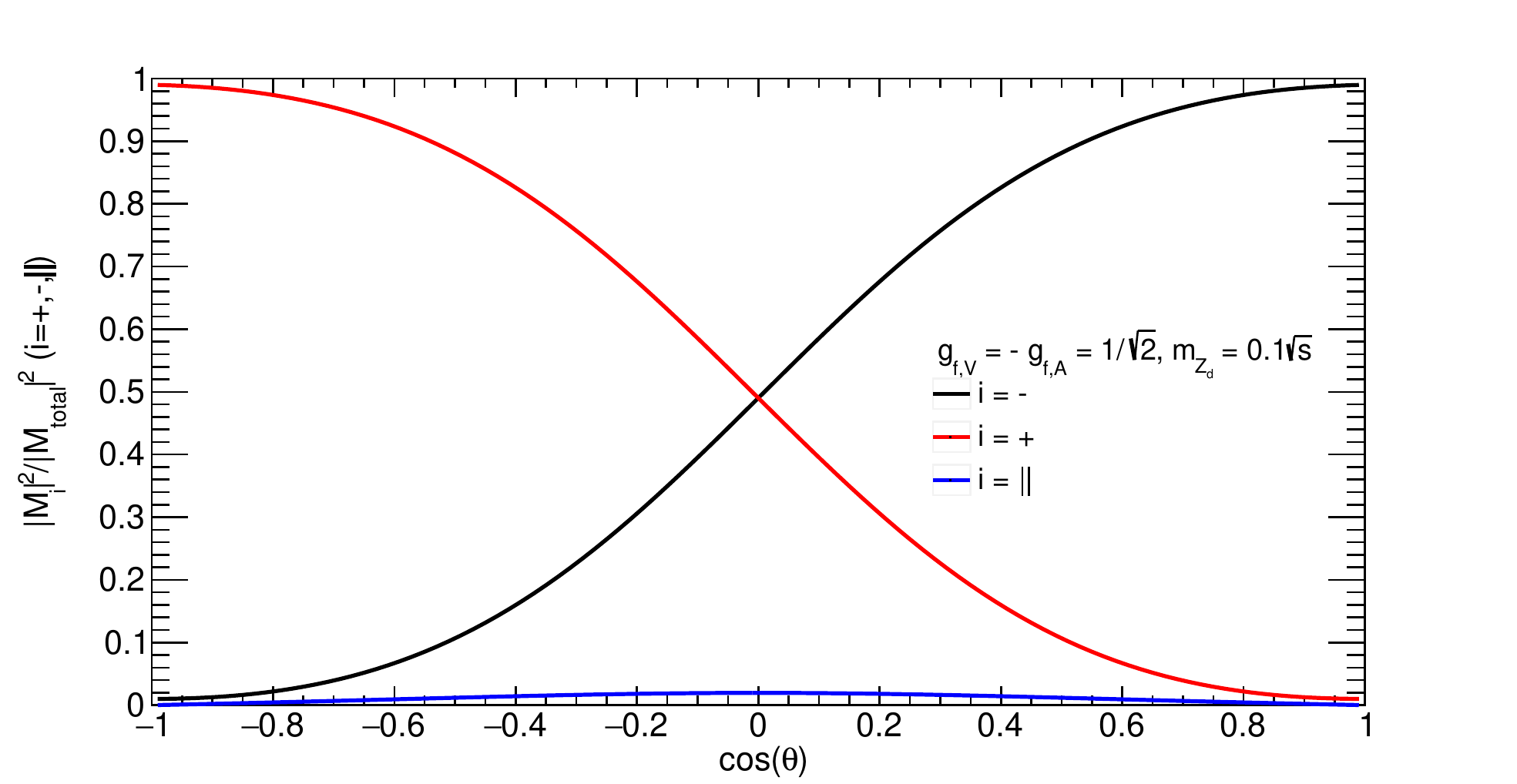}
		\includegraphics[width=12cm]{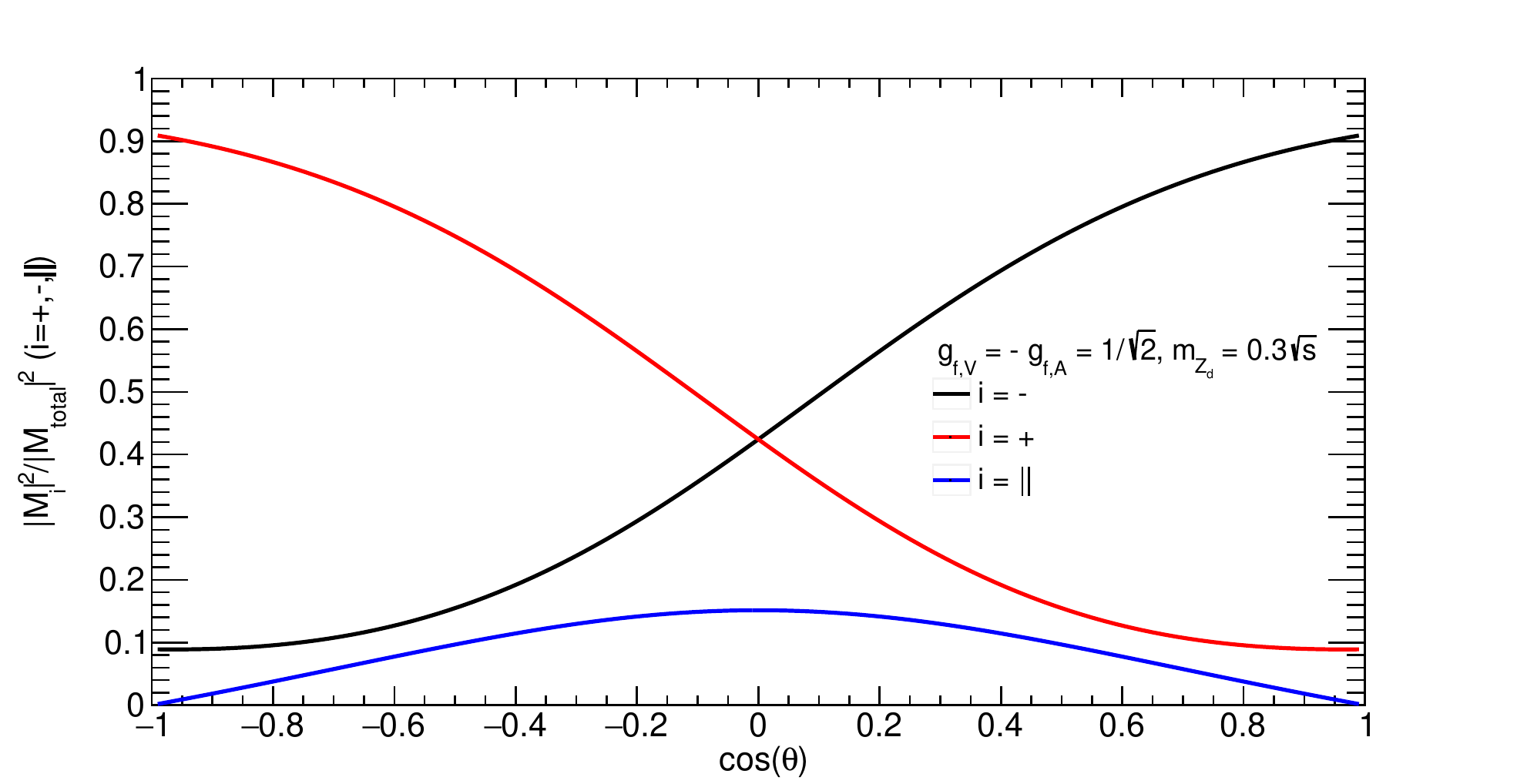}
		\includegraphics[width=12cm]{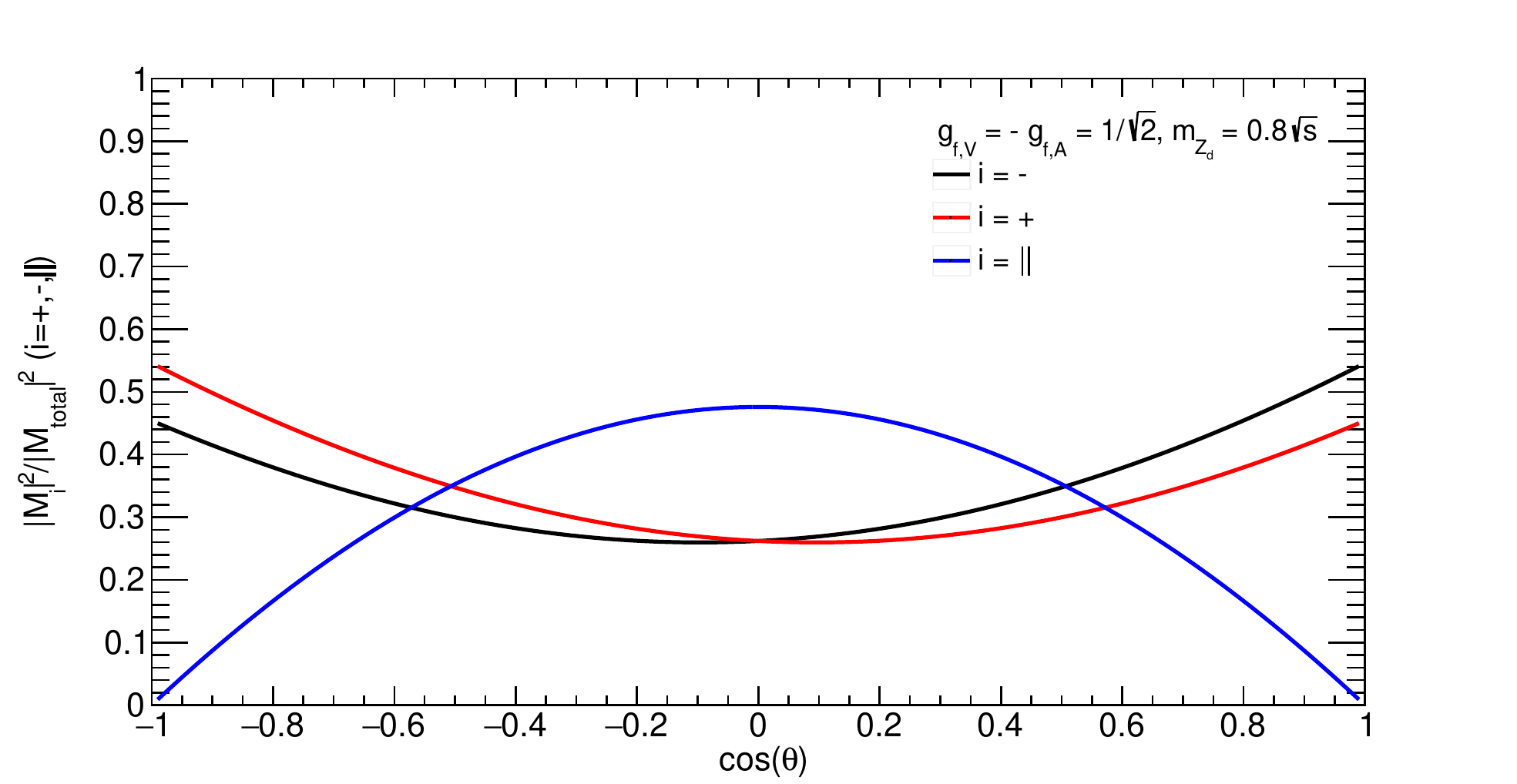}	
		\caption{The fraction of matrix element square for helicity $+1$, $-1$, and longitudinal dark boson final state. We take the $V-A$ case with $g_{f,V}=-g_{f,A}=1/\sqrt{2}$. The ratio $m_{Z_d}/\sqrt{s}$ is taken to be $0.1$, $0.3$, and $0.8$ on upper, middle, and lower panels, respectively.    }
		\label{fig:amp_square}
	\end{center}
\end{figure}
where $\omega$ is the photon energy, $E_{Z_d}=\sqrt{\omega^2+m_{Z_d}^2}$, and $s=(p_1+p_2)^2=4E^2$. Using energy and momentum conservation, we have $E_{Z_d}=E+m_{Z_d}^2/(2\sqrt{s})$ and
$\omega=E-m_{Z_d}^2/(2\sqrt{s})$. Let us denote the amplitude for each polarization as ${\cal M}_+$, ${\cal M}_-$, and ${\cal M}_{\parallel}$ for right-handed, left-handed, and longitudinal polarized dark boson final state, respectively. We have
\begin{eqnarray}
\vert \bar{{\cal M}}\vert ^2_+ &=& \frac{8\pi^2\alpha^2\varepsilon^2}{(t-m_e^2)(u-m_e^2)}\bigg[(1+\cos^2\theta)(s^2+m_{Z_d}^4)+\rho\cos\theta (s-m_{Z_d}^2)^2\bigg], \nonumber \\
\vert \bar{{\cal M}}\vert ^2_- &=& \frac{8\pi^2\alpha^2\varepsilon^2}{(t-m_e^2)(u-m_e^2)}\bigg[(1+\cos^2\theta)(s^2+m_{Z_d}^4)-\rho\cos\theta (s-m_{Z_d}^2)^2\bigg], \nonumber \\
\vert \bar{{\cal M}}\vert ^2_{\parallel}&=&\frac{8\pi^2\alpha^2\varepsilon^2}
{(t-m_e^2)(u-m_e^2)}(4m_{Z_d}^2 s\sin^2 \theta),
\label{eq:amp_square}
\end{eqnarray} 
where the normalization $g_{f,V}^2+g_{f,A}^2=1$ has been taken and 
$\rho=4g_{f,V}g_{f,A}$. The absolute value of $\rho$ essentially describes the degree of parity violation.  
It is shown that $\vert \bar{{\cal M}}\vert ^2_{\parallel}$ is suppressed by $m_{Z_d}^2/s$ compared to $\vert \bar{{\cal M}}\vert ^2_{\pm}$. In Fig.~\ref{fig:amp_square}, we present the fraction of matrix element square as a function of $\cos\theta$ for each helicity state of $Z_d$. We take the $V-A$ case with $g_{f,V}=-g_{f,A}=1/\sqrt{2}$ for illustration. The upper, middle, and lower panels correspond to $m_{Z_d}/\sqrt{s}=0.1$, $0.3$, and $0.8$, respectively. For the first two cases, one can see that the longitudinal fraction is no more than $10\%$. In addition, $\vert \bar{{\cal M}}\vert ^2_-$ dominates the forward direction ($0\leq \cos \theta \leq 1$) while $\vert \bar{{\cal M}}\vert ^2_+$ dominates the backward direction. For the third case, the longitudinal fraction is non-negligible and the fractions for helicity $+1$ and $-1$ states are almost identical due to the suppression of $\rho$ dependent terms, i.e., the forward-backward asymmetry approaches to zero in the limit $s\to m_{Z_d}^2$. 
\begin{figure}[htbp]
	\begin{center}
		\includegraphics[width=12cm]{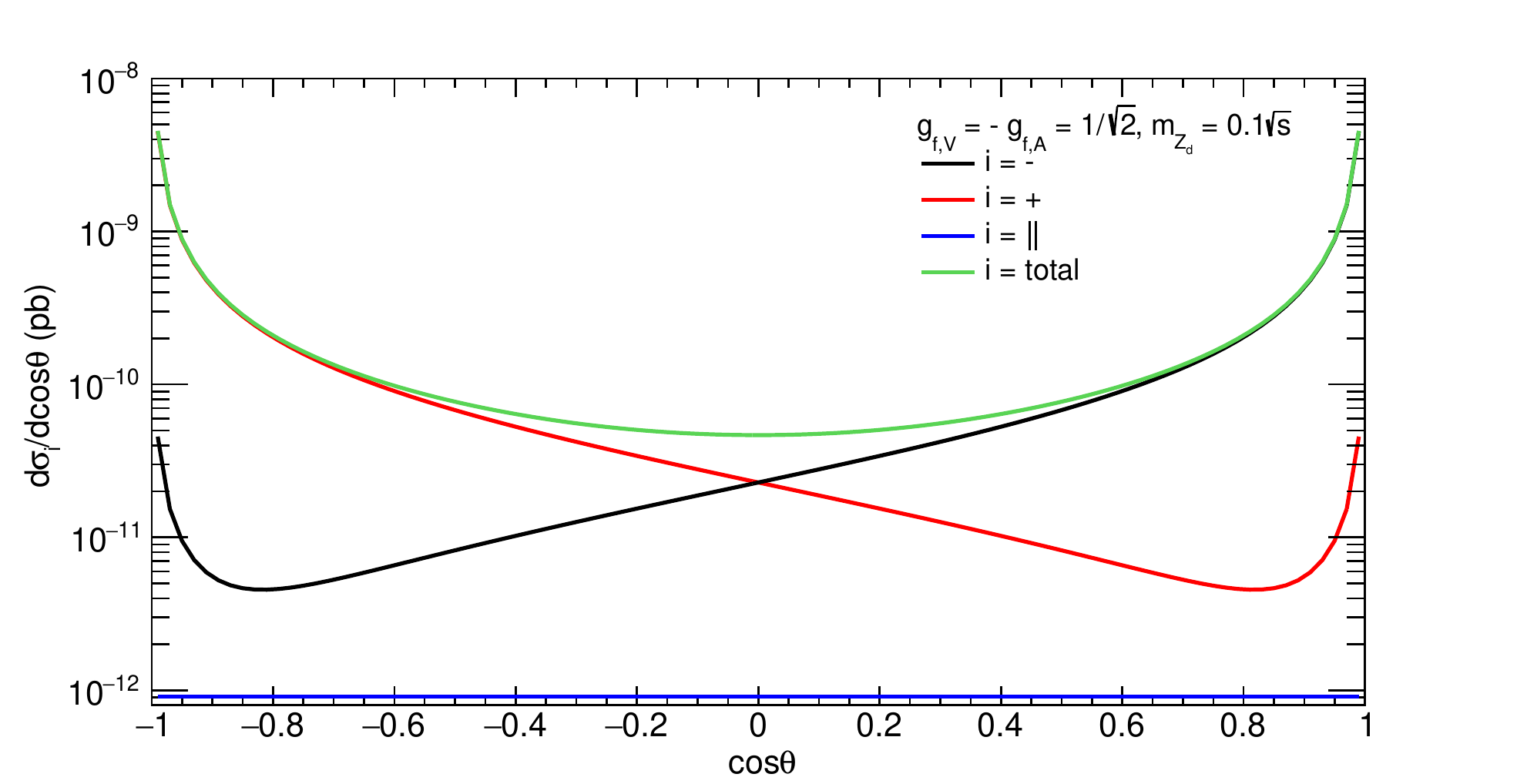}
		\includegraphics[width=12cm]{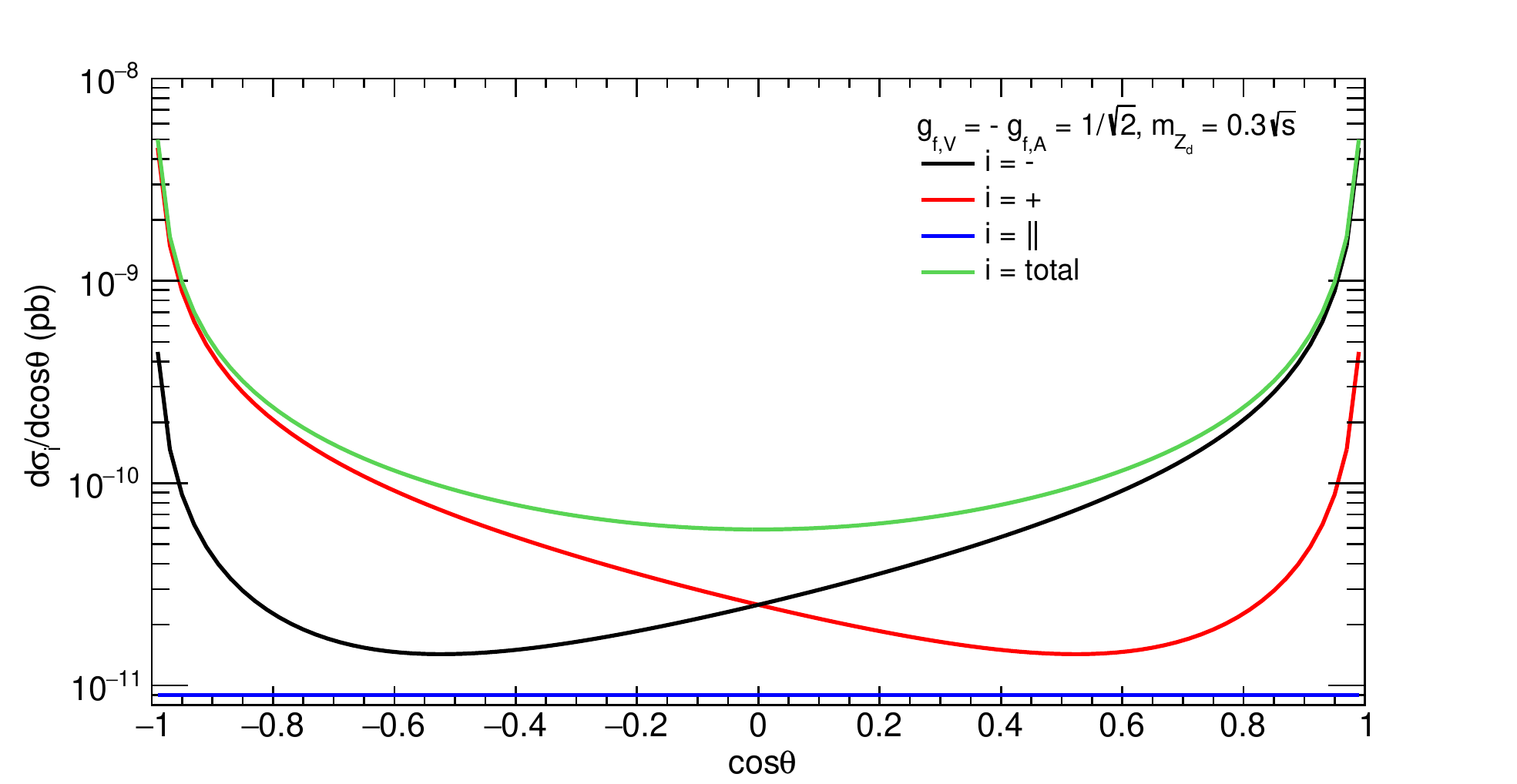}
		\includegraphics[width=12cm]{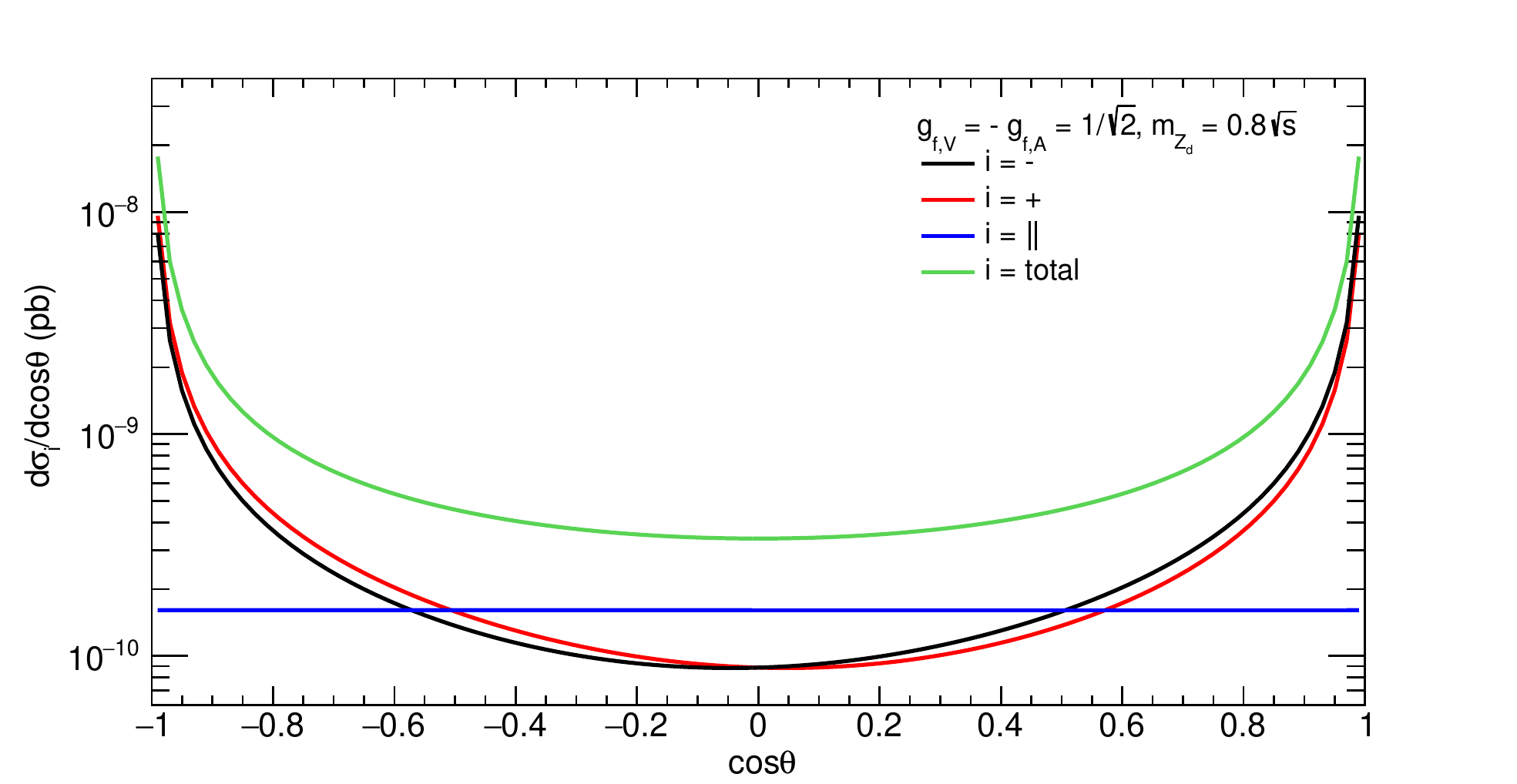}	
		\caption{The polarized differential cross section for $e^+  e^-\to Z_d  \gamma$ with $\varepsilon=7\times 10^{-4}$ at Belle II for $m_{Z_d}/\sqrt{s}=0.1, \ 0.3,$ and $0.8$, respectively.   }
		\label{fig:cross_section}
	\end{center}
\end{figure}

\begin{figure}[htbp]
	\begin{center}
		\includegraphics[width=12cm]{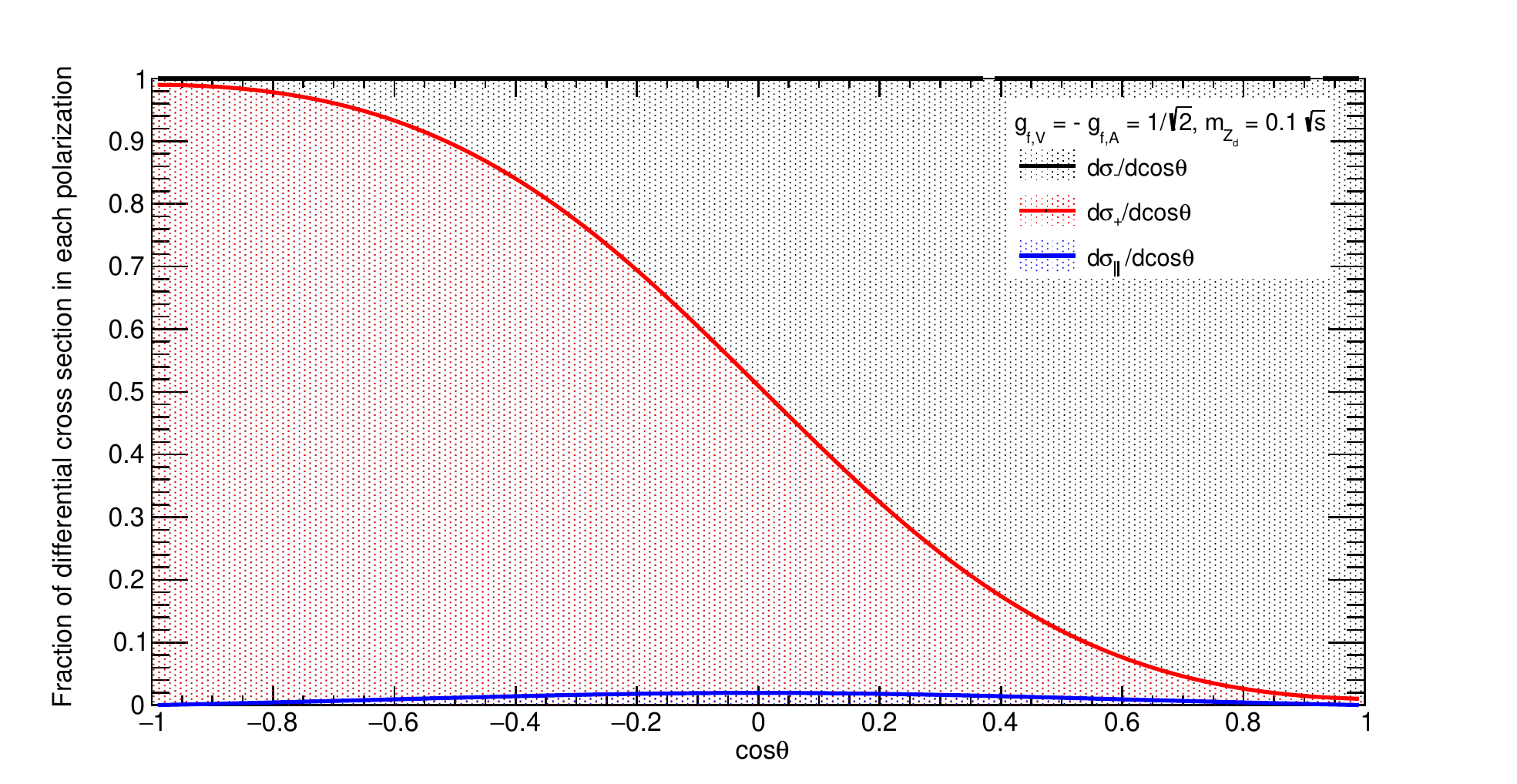}
		\includegraphics[width=12cm]{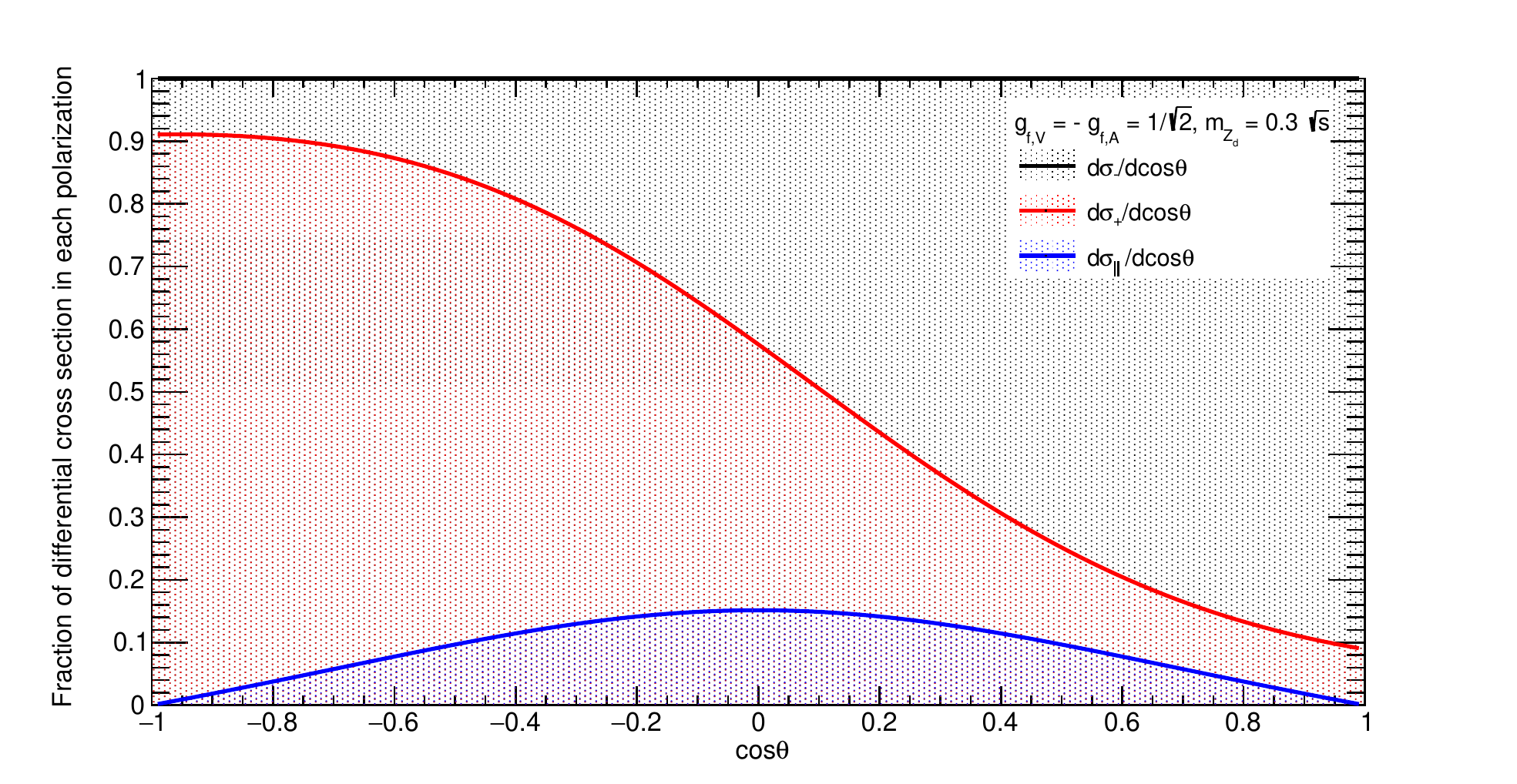}
		\includegraphics[width=12cm]{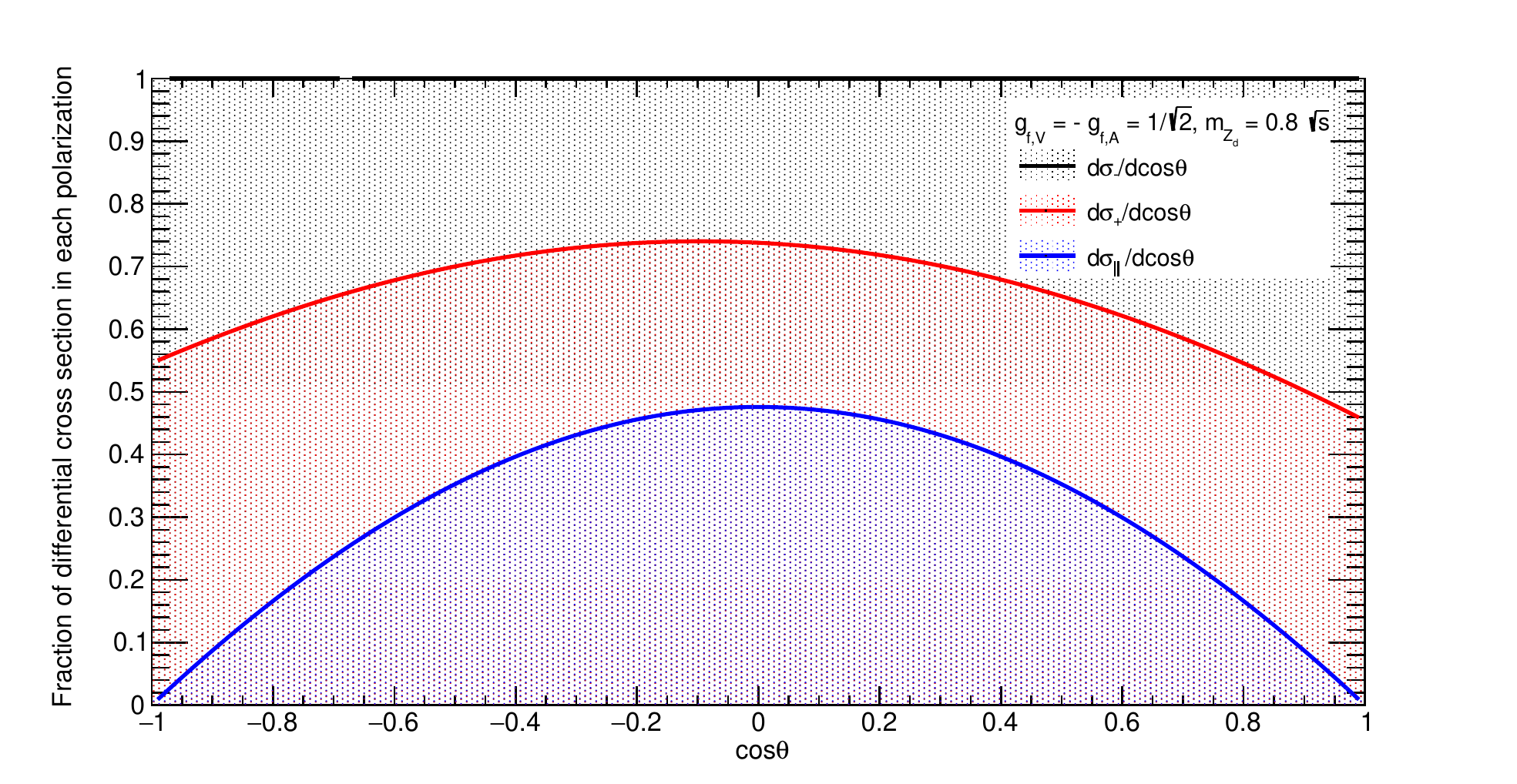}	
		\caption{The fraction of differential cross section in each $Z_d$ polarization as represented by regions of different colors. The range for each color adds up to unity.  }
		\label{fig:area}
	\end{center}
\end{figure}

The polarized differential cross section is readily calculated with 
\begin{eqnarray}
\frac{d\sigma_i}{d\cos\theta}=\frac{1}{32\pi s}(1-\frac{m_{Z_d}^2}{s})\vert {\cal M} \vert ^2_i,
\label{eq:cross_section}
\end{eqnarray}
with $i=+, \ -$, and $\parallel$. To check our result, we take $\varepsilon=1$ and sum over contributions from all polarizations. In the limit $s\gg m_e^2,m_{Z_d}^2$, our result approaches to the differential cross section of $e^+e^-\to \gamma\gamma$. The results for polarized differential cross section are shown in Fig.~\ref{fig:cross_section}. In this calculation we take $\varepsilon=7\times 10^{-4}$ for $m_{Z_d}/\sqrt{s}=0.1$, $0.3$, and $0.8$, respectively, for illustrations. This $\varepsilon$ value is reachable by Belle II with $500$ fb$^{-1}$ luminosity for $m_{Z_d}$ around $1$ GeV~\cite{Kou:2018nap}. Similar to the case of amplitude square, the longitudinal polarized contribution is suppressed for small $m_{Z_d}/\sqrt{s}$. We present in Fig.~\ref{fig:area} the fraction of differential cross section in each $Z_d$ polarization. Each fraction is represented by a region of specific color.   

\section{Probing parity violation effects in $e^+e^-\to Z_d\gamma\to Z_d\mu^+\mu^-$ }
Now we consider the angular distributions of $Z_d$ decays. Through mixing with SM gauge bosons, $Z_d$ can decay to SM leptons with the polarized differential decay rate 
\begin{eqnarray}
\frac{d\Gamma^i_{l^+l^-}}{d\cos\theta_{d}}=\dfrac{\beta}{32\pi m_{Z_d}}\vert {\cal M}(Z_{d}^{(i)}\rightarrow l^{+}l^{-} ) \vert ^2,
\label{eq:diff_decay_rate_gen}
\end{eqnarray}
with $i=+, \ -$, and $\parallel$, $\beta=p_l/E_l=\sqrt{1-4m_l^2/m_{Z_d}^2}$, and $\theta_{d}$  the angle between $l^{-}$ direction in the $Z_d$ rest frame and the $Z_d$ direction in $e^+e^-$ CM frame. Thus we obtain
\begin{eqnarray}
\dfrac{d\Gamma^{+}_{l^+l^-}}{d\cos\theta_{d}} &=& \dfrac{\alpha\varepsilon^{2}\beta}{2m_{Z_d}}\left[2g_{l,V}^2 m_{l}^{2} +(1+\cos^{2}\theta_{d})p_{l}^{2}+\rho \cos\theta_{d}E_{l}p_{l}\right] , \nonumber \\
\dfrac{d\Gamma^{-}_{l^+l^-}}{d\cos\theta_{d}} &=& \dfrac{\alpha\varepsilon^{2}\beta}{2m_{Z_d}}\left[2g_{l,V}^2 m_{l}^{2} +(1+\cos^{2}\theta_{d})p_{l}^{2}-\rho \cos\theta_{d}E_{l}p_{l}\right] , \nonumber \\
\dfrac{d\Gamma^{\parallel}_{l^+l^-}}{d\cos\theta_{d}} &=& \dfrac{\alpha\varepsilon^{2}\beta}{m_{Z_d}}\left[g_{l,V}^2 m_{l}^{2} +\sin^{2}\theta_{d} p_{l}^{2}\right].
\label{eq:diff_decay_rate} 
\end{eqnarray}     
Given $g_{l,V}^2+g_{l,A}^2=1$ and $\rho=4g_{l,V}g_{l,A}$, we have $g_{l,V}^2=(1+\sqrt{1-\rho^{2}/4})/2$ for $\vert g_{l,V} \vert \geq \vert g_{l,A} \vert$, while 
$g_{l,V}^2=(1-\sqrt{1-\rho^{2}/4})/2$ for $\vert g_{l,V} \vert \leq \vert g_{l,A} \vert$. 
\begin{figure}[htbp]
	\begin{center}
		\includegraphics[width=12cm]{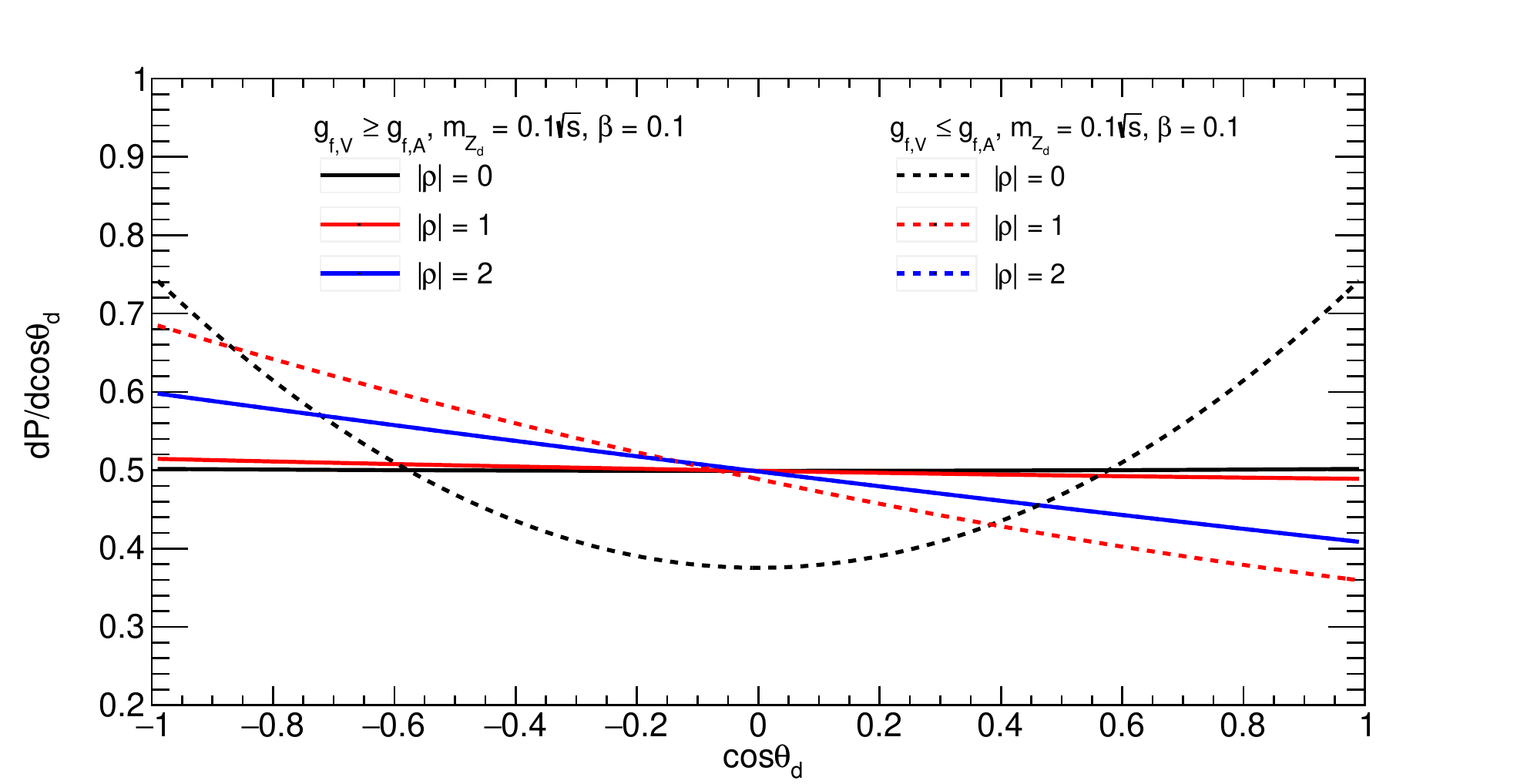}
		\includegraphics[width=12cm]{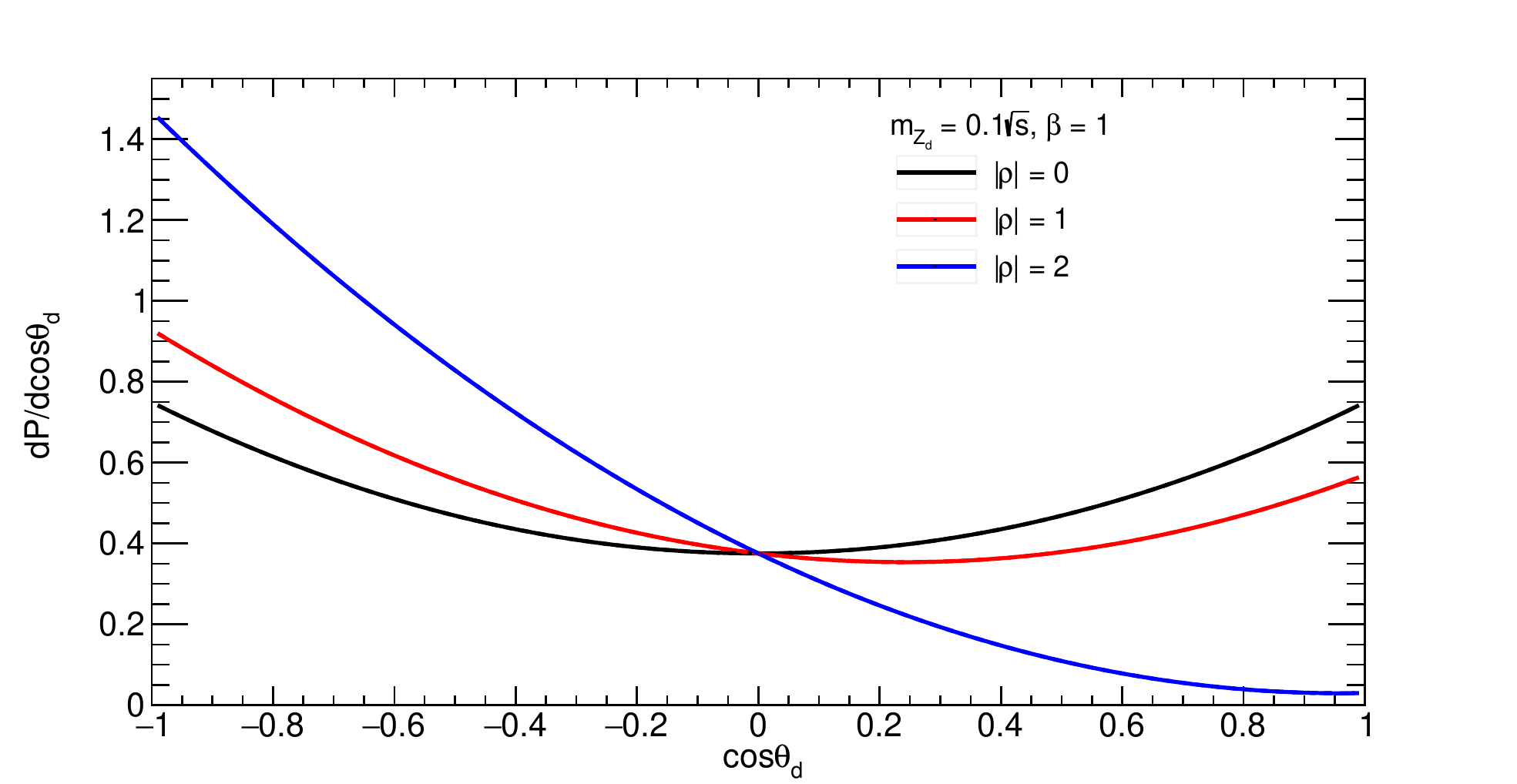}
		\includegraphics[width=12cm]{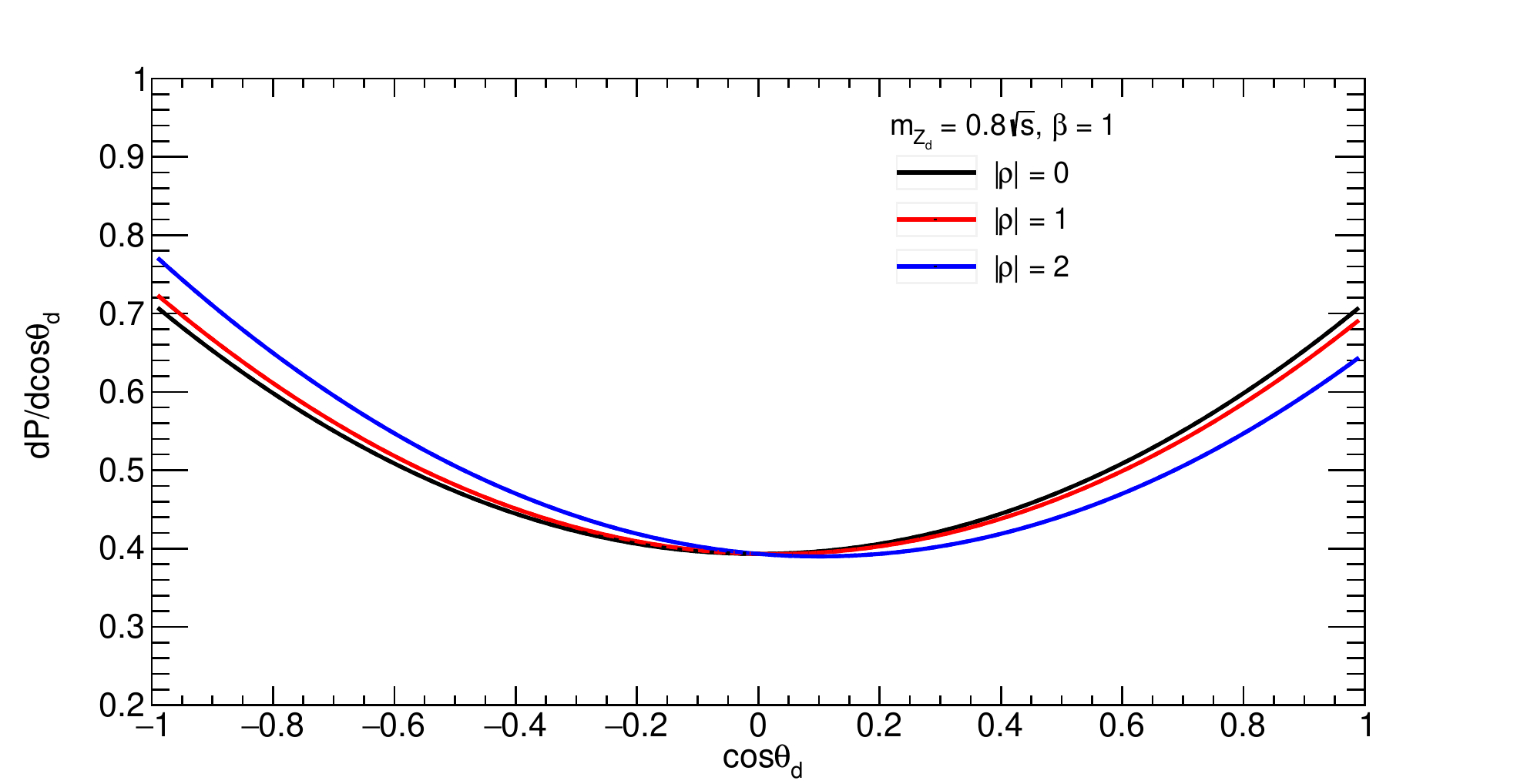}	
		\caption{The angular distribution of $l^-$ from decays of $Z_d$ produced in the backward direction by $e^+e^-\to Z_d\gamma$ for $\vert \rho \vert=0$, $1$, and $2$, respectively.  The upper panel is for relativistic $Z_d$ with $m_{Z_d}=0.1\sqrt{s}$ and non-relativistic lepton with $\beta=0.1$. The middle panel corresponds to $(\beta,m_{Z_d}/\sqrt{s})=(1,0.1)$ while the lower panel corresponds to $(\beta,m_{Z_d}/\sqrt{s})=(1,0.8)$. 
		}
		\label{fig:prob}
	\end{center}
\end{figure}
The double differential distribution of final-state leptons is given by 
\begin{eqnarray}
\dfrac{d^2P}{d\kappa d\xi}=\dfrac{1}{\sigma_T\cdot \Gamma_{l^+l^-}}\sum_i  \left( \dfrac{d\sigma^i}{d\cos\theta}\right)\cdot \left(\dfrac{d\Gamma^{i}_{l^+l^-}}{d\cos\theta_{d}}\right),
\label{double_distribution}
\end{eqnarray}
 with $\kappa=\cos\theta$, $\xi=\cos\theta_d$, $\Gamma_{l^+l^-}$ the unpolarized $Z_d\to l^+l^-$ decay width, and $\sigma_T$ the total $Z_d$ production cross section. We first observe that the double differential distribution $d^2P/d\kappa d\xi$ only depends on $\rho^0$ and $\rho^2$. Secondly, the $\rho^2$ dependent term in the double distribution is given by $\rho^2\beta(1-m_{Z_d}^2/s)^2J/(1-\kappa^2)$ with $J=\kappa\xi$. The sign of this contribution is determined by the sign of $J$.  This contribution vanishes at the dark boson production threshold, $s=m_{Z_d}^2$, or at the threshold for $Z_d$ decaying into the lepton pair, i.e., $\beta=0$. It also vanishes if either $\kappa$ or $\xi$ is integrated to the full range. 
 
To analyze the double distribution, we integrate $\kappa$ from $-1$ to $0$, i.e., we consider leptonic decays of $Z_d$ produced in the backward direction. The forward-backward asymmetry of $l^-$ in its helicity angle is presented in  Fig.~\ref{fig:prob}. In three panels we present the angular distribution of $l^-$ for $\vert \rho \vert=0$, $1$, and $2$, respectively. The case $\vert \rho\vert=0$ implies either $g_{l,A}=0$ or $g_{l,V}=0$ with the former corresponding to the dark-photon scenario, while $\vert \rho\vert=2$ corresponds to either $V-A$ or $V+A$ cases. On the upper panel, we separate results into $\vert g_{l,V}\vert \geq \vert g_{l,A}\vert$ and $\vert g_{l,V}\vert \leq \vert g_{l,A}\vert$. These two cases coincide in the middle and lower panels with $\beta=1$, i.e., $m_l\to 0$. The asymmetries in the upper and lower panels are small either due to a small $\beta$ or to the suppression in $(1-m_{Z_d}^2/s)^2$ with $m_{Z_d}/\sqrt{s}=0.8$. Significant asymmetry is seen in the middle panel with large $\vert \rho\vert$. 
In general, we may define the following asymmetry parameter
 \begin{eqnarray}
\mathcal{A}_{\rm PN}\equiv \dfrac{S(\kappa\cdot \xi>0)-S(\kappa\cdot \xi<0)}{S(\kappa\cdot \xi<0)+S(\kappa\cdot \xi>0)},
\label{eq:asymmetry}
\end{eqnarray}
where the subscript PN indicates that $\mathcal{A}_{\rm PN}$ describes the difference in signal event rate as $\kappa\xi$ reverses its sign.
In limits of $\beta\to 1$ and $m_{Z_d}\ll \sqrt{s}$, we have  
 \begin{eqnarray}
 \mathcal{A}_{\rm PN}=\frac{3}{4}\left(\frac{\rho^2}{4}\right)
 \dfrac{-\ln \left(1-\kappa_{m}^2\right)}{\ln \left( \frac{1+\kappa_{m}}{1-\kappa_{m}}\right)-\kappa_{m}},
 \label{eq:APN_limit}
 \end{eqnarray}  
where $\kappa_m$ is the maximum of $\kappa$. The minimum of $\kappa$ is assumed to be $-\kappa_m$. It is found that $\mathcal{A}_{\rm PN}$ is not very sensitive to $\kappa_m$. $\mathcal{A}_{\rm PN}=0.64
\times (\rho^2/4)$ for $\kappa_m=0.95$, and $0.55\times (\rho^2/4)$ for $\kappa_m=0.80$. 
In the next section, we shall calculate $\mathcal{A}_{\rm PN}$ and its associated uncertainty with acceptance cuts in Belle II detector.

\section{The prospect of measuring parity-violating dark boson interactions in Belle II}

In this section, we discuss the search for dark vector boson and the possible measurement of parity violation parameter $\rho$ through determining $\mathcal{A}_{\rm PN}$ in Belle II detector. We shall begin by considering the detector acceptance of Belle II and compare our sensitivity estimation for the dark photon search through $e^+e^-\to A^{\prime} \gamma$ with $A^{\prime}\to \mu^+\mu^-$ with BaBar result at $514$ fb$^{-1}$ 
and the projected sensitivity of Belle II at $500$ fb$^{-1}$. These comparisons are important for validating our approach. Next we consider Belle II at the full integrated luminosity $50$ ab$^{-1}$ and extend our discussions to the dark boson scenario with a non-vanishing $\rho$. 
\subsection{Sensitivity for the dark photon search at Belle II} 
To illustrate our points in previous sections, we take the dark photon mass as $0.5$ GeV and $2$ GeV, respectively as benchmark values.  
These two mass values satisfy $m_{Z_d}/\sqrt{s}\ll 1$ so that the dark photons are produced in transversely polarized states. Hence for the generalization to dark boson scenario in the next subsection, we shall see that the asymmetry parameter ${\mathcal A}_{\rm PN}$ will be significant.

Let us begin by taking $m_{Z_d}=0.5$ GeV with $\rho=0$. In this case, the branching ratio for $Z_d\to \mu^+\mu^-$ is about $40\%$ from the measurement of $R=\sigma (e^+e^-\to {\rm hadrons})/\sigma(e^+e^-\to \mu^+\mu^-)$~\cite{Batell:2009yf,Tanabashi:2018oca}. The Belle II calorimeter angular coverage is 
$12.4^{\circ}\leq \theta_{\gamma}^{\rm lab} \leq 155.1^{\circ}$~\cite{Adachi:2018qme}, which detects final-state photon
in the rapidity range $-1.51\leq \eta_{\gamma}^{\rm lab}\leq 2.22$. Since the boost velocity from the laboratory frame to CM frame is $\beta_{\rm CM}=(E_{e^-}-E_{e^+}) /(E_{e^-}+E_{e^+})=3/11$, the photon rapidity in the CM frame is given by $\eta_{\gamma}^{\rm CM}=\eta_{\gamma}^{\rm lab}+\ln((1-\beta_{\rm CM})/(1+\beta_{\rm CM}))/2$. Hence $-1.79\leq \eta_{\gamma}^{\rm CM}\leq 1.94$. Furthermore the angular coverage of $K_L$-Muon detector~\cite{Adachi:2018qme} is $25^{\circ}\leq \theta_{\mu^{\pm}}^{\rm lab} \leq 150^{\circ}$. This leads to the muon rapidity range $-1.60\leq \eta_{\mu^{\pm}}^{\rm CM}\leq 1.23$ in the CM frame. Since the signal mass resolution is between $1.5$ MeV and $8$ MeV in BaBar analysis~\cite{Lees:2014xha}, we take it to be $5$ MeV for our sensitivity estimation, i.e., the signal and background events are calculated within a $5$ MeV $\mu^{+}\mu^{-}$ invariant mass window in the vicinity of assumed $Z_{d}$ mass.    

We note that the rapidity cuts preserve $S(\kappa\cdot \xi>0)=S(\kappa\cdot \xi<0)$ for $\rho=0$. 
Using CalcHEP~\cite{Belyaev:2012qa}, we find that the signal $e^+e^-\to \gamma Z_d$ with $Z_d\to \mu^+\mu^-$ has the cross section $1.84\cdot 10^3 \cdot \varepsilon^2$ pb by considering both the rapidity cuts and the $40\%$ 
$Z_{d} \rightarrow \mu^{+}\mu^{-}$ branching ratio, and the cross section for QED background process $e^+e^-\to \gamma \mu^+\mu^-$ with the same acceptance cut is $7.76\cdot 10^{-2}$ pb. We note that the above parametrization for signal cross section is valid only for $\varepsilon< 0.3$ such that the $\varepsilon$-dependent  $Z_d$ width is less than $10\%$ of the signal mass resolution. With $500 \ {\rm fb}^{-1}$ of integrated luminosity, the Belle II $90\%$ C.L. sensitivity to $\varepsilon$ is estimated
by the following $\chi^2$ function
 \begin{eqnarray}
\chi^2=2\left( n\ln(\frac{n}{w})+w-n \right),
 \label{eq:chisquare_1}
 \end{eqnarray}          
where $n$ is the observed event number while $w$ is the expected event number. With $n=S+B=(1.84\cdot 10^3 \cdot \varepsilon^2+7.76\cdot 10^{-2}) \ {\rm pb}\cdot 500 \ {\rm fb}^{-1}$, $w=B=7.76\cdot 10^{-2} \ {\rm pb} \cdot 500 \ {\rm fb}^{-1}$, and $\chi^2=(1.645)^2$ for $90\%$ C.L. sensitivity\footnote{In principle one should also consider the interference between signal and background amplitudes for calculating $S+B$. On the other hand, it can be shown that such contribution scales as $\varepsilon^2$, and for the current case the interference part of the cross section is $\sim -1.8\cdot \varepsilon^2$ pb, which is negligible.}, we
obtain $\varepsilon=5.9\cdot 10^{-4}$, which is consistent with the sensitivity $\varepsilon=5.6\cdot 10^{-4}$ given in Belle II physics book for the visible modes $Z_d\to e^+ e^-, \  \mu^+\mu^-$~\cite{Kou:2018nap}. The latter is also comparable to the constraint from BaBar search via visible modes at $514$ fb$^{-1}$~\cite{Lees:2014xha}.

We next take $m_{Z_d}=2$ GeV with $\rho=0$. In this case, the branching ratio of $Z_d\to \mu^+\mu^-$ is about $24\%$~\cite{Tanabashi:2018oca}. Hence the signal cross section is around $1.11\cdot 10^3
\cdot \varepsilon^2$ pb while the background cross section is $2.54\cdot 10^{-2}$ pb. Following Eq.~(\ref{eq:chisquare_1}), we obtain Belle II $90\%$ C.L. sensitivity to $\varepsilon$ as $\varepsilon=5.8\cdot 10^{-4}$, which is also not much different from $6.6\cdot 10^{-4}$ given by Belle II physics book for the visible modes $Z_d\to e^+ e^-, \  \mu^+\mu^-$~\cite{Kou:2018nap}.

\subsection{Probing the parity violation effects} 

\subsubsection{Enhancement on the detection significance} 
In the case of non-vanishing $\rho$, we modify the $\chi^2$ function in Eq.~(\ref{eq:chisquare_1}) into 
\begin{eqnarray}
\chi^2=2\left( n_{a}\ln(\frac{n_{a}}{w_{a}})+w_{a}-n_{a} \right)+2\left( n_{b}\ln(\frac{n_{b}}{w_{b}})+w_{b}-n_{b} \right),
 \label{eq:chisquare_2}
 \end{eqnarray}      
where $n_a$ ($w_a$) and $n_b$ ($w_b$) are observed (expected) event numbers in $\kappa \cdot \xi>0$ and $\kappa \cdot \xi<0$ bins, respectively.
By considering separate event bins, the dark boson detection significance is expected to be improved. With $n_{a,b}=S_{a,b}+B_{a,b}$ and $w_{a,b}=B_{a,b}$, we can show that 
\begin{eqnarray}
\chi^2=\frac{S^2}{B}(1+\mathcal{A}^2_{\rm PN}),
\label{eq:sig_improve}
\end{eqnarray} 
with the assumption $S_{a,b}\ll B_{a,b}$ and the identity $B_a=B_b=B/2$. We note that $S_a$ and $S_b$ are the event excess in $\kappa\cdot \xi>0$ and $\kappa\cdot \xi<0$ event bins, respectively. 
It is clear that the detection significance increases from $[S/\sqrt{B}]\cdot \sigma$ 
to $[S\sqrt{(1+\mathcal{A}^2_{\rm PN})}/\sqrt{B}]\cdot \sigma$ by considering separate event bins.

\subsubsection{Measurement of ${\mathcal A}_{\rm PN}$}
Following Eq.~(\ref{eq:asymmetry}), we have ${\mathcal A}_{\rm PN}=(S_a-S_b)/(S_a+S_b)$ with the statistical uncertainties of both numerator and denominator being $\sqrt{B}$. Hence the uncertainty of 
${\mathcal A}_{\rm PN}$ is given by $\sigma_{{\mathcal A}_{\rm PN}}=\sqrt{1+{\mathcal A}_{\rm PN}^2}(\sqrt{B}/S)$.
We note that the asymmetry parameter ${\mathcal A}_{\rm PN}$ as defined by Eq.~(\ref{eq:asymmetry}) is actually independent of the integrated luminosity. It can be calculated by the scattering cross section with appropriate kinematic cuts imposed. 

\subsubsection{Numerical results}
Let us assume the scenario that one achieves a 5 standard deviation detection of dark boson at the designed integrated luminosity $50$ ab$^{-1}$ based upon counting the overall event excess, i.e., $S= 5\sqrt{B}$ according to 
Eq.~(\ref{eq:chisquare_1}). On the other hand, the event excess $S_a$ and $S_b$ for each event bin depends on the asymmetry parameter ${\mathcal A}_{\rm PN}$. It is to be noted that we shall fix the total event excess $S\equiv S_a+S_b$ regardless of the $\rho$ value. Hence $\sigma(e^+e^-\to Z_d\gamma)\cdot {\rm Br}(Z_d\to \mu^+\mu^-)$ is fixed. Since $\sigma(e^+e^-\to Z_d\gamma)\propto \varepsilon^2$ while ${\rm Br}(Z_d\to \mu^+\mu^-)$ depends on $\rho$, the $\varepsilon$ value extracted from $S= 5\sqrt{B}$ at $50$ ab$^{-1}$ also depends on $\rho$.      

To proceed with our numerical analysis, we first calculate ${\mathcal A}_{\rm PN}$ and its uncertainty as a function of the ratio $m_{Z_d}/\sqrt{s}$ for a given $\rho$.
\begin{figure}[htbp]
	\begin{center}
		\includegraphics[width=.50\columnwidth]{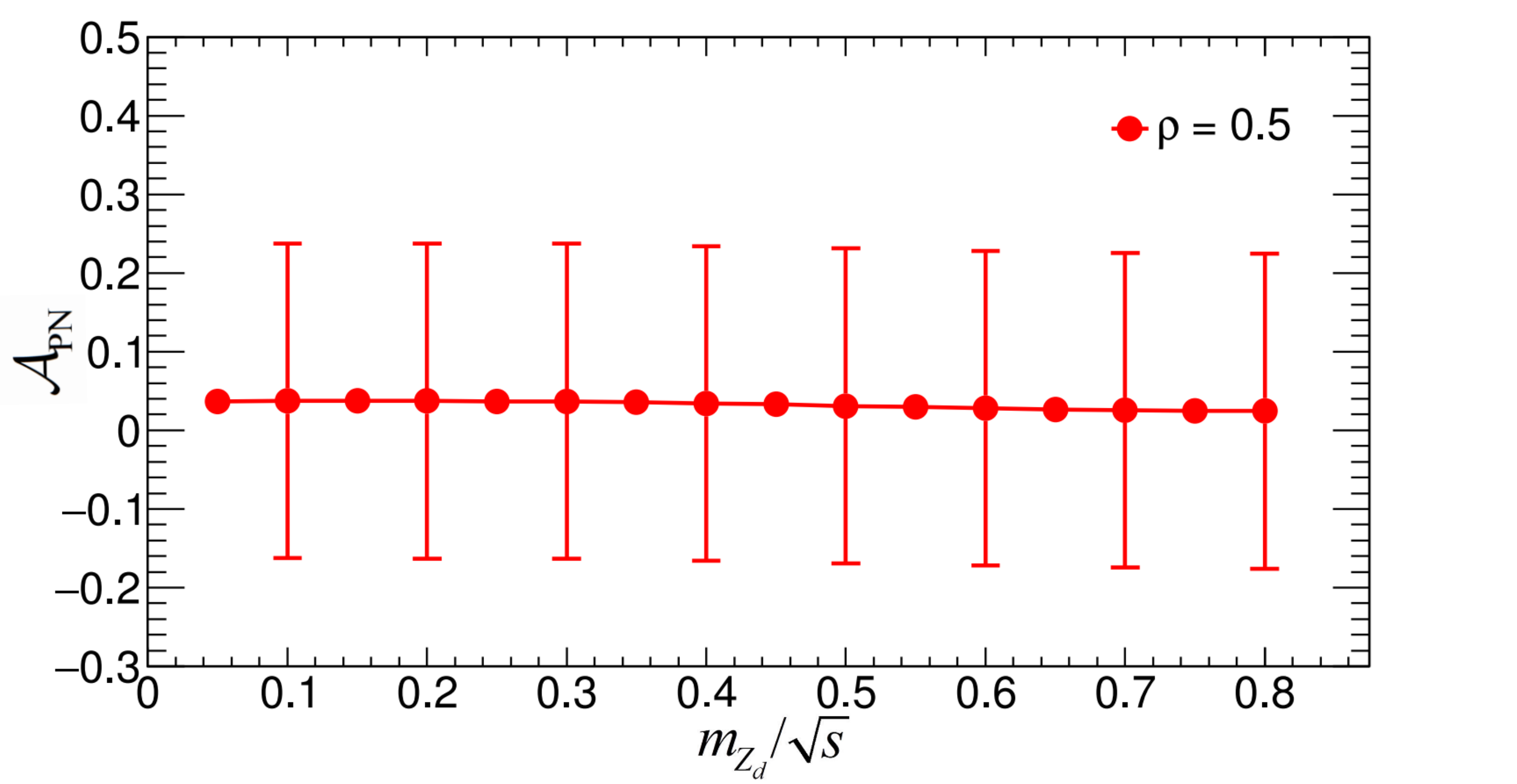}%
		\includegraphics[width=.50\columnwidth]{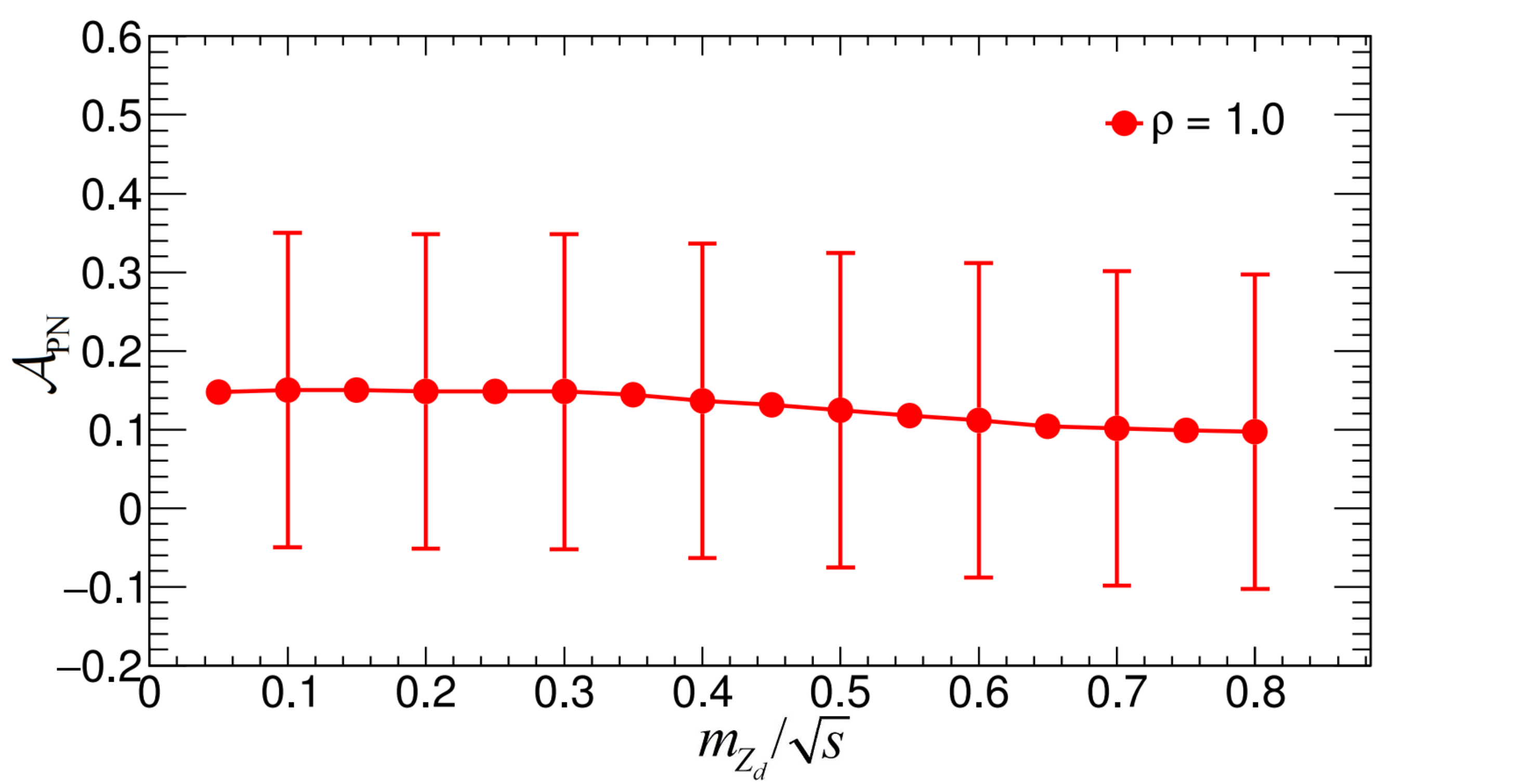} \\%
		\includegraphics[width=.50\columnwidth]{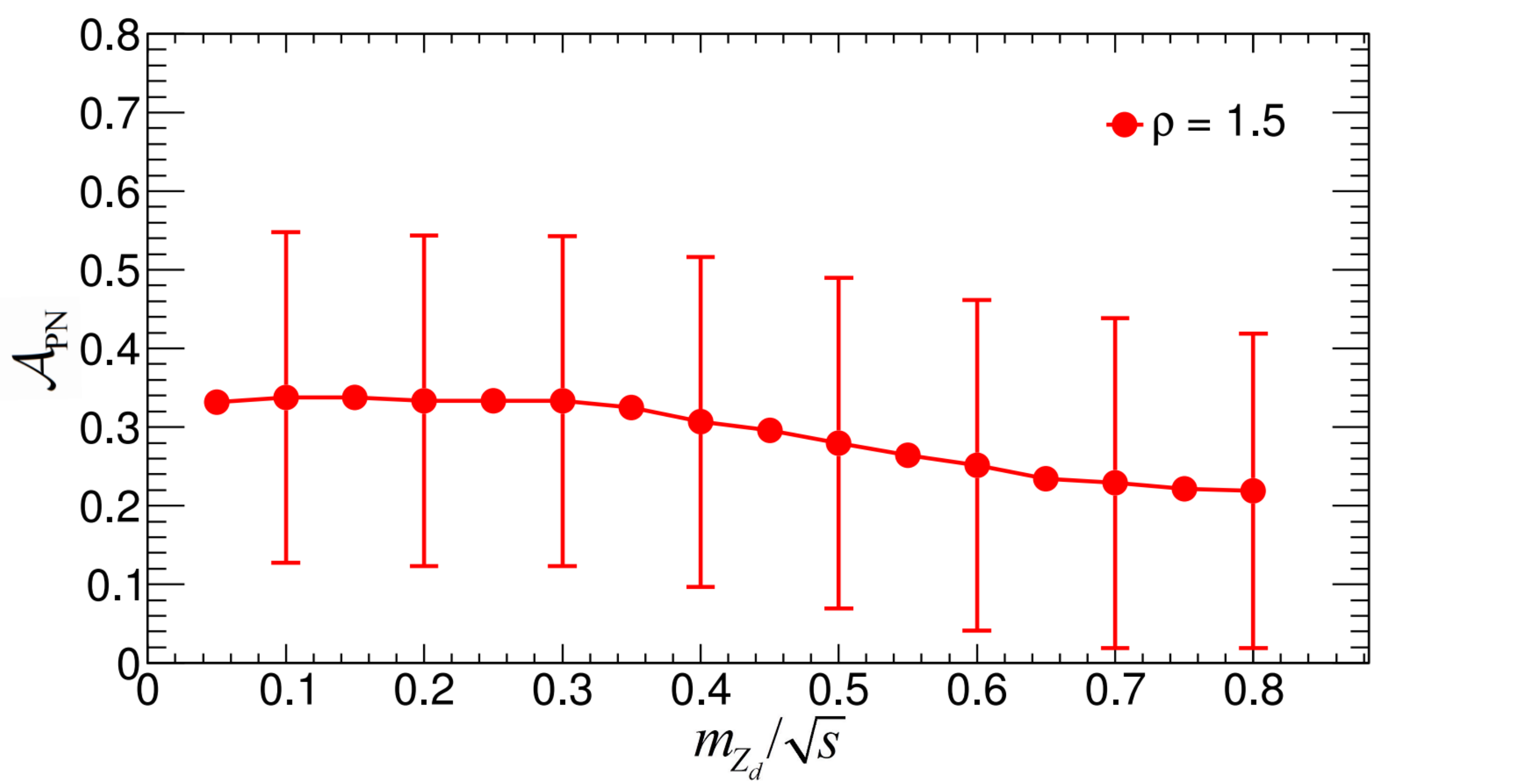}%
		\includegraphics[width=.50\columnwidth]{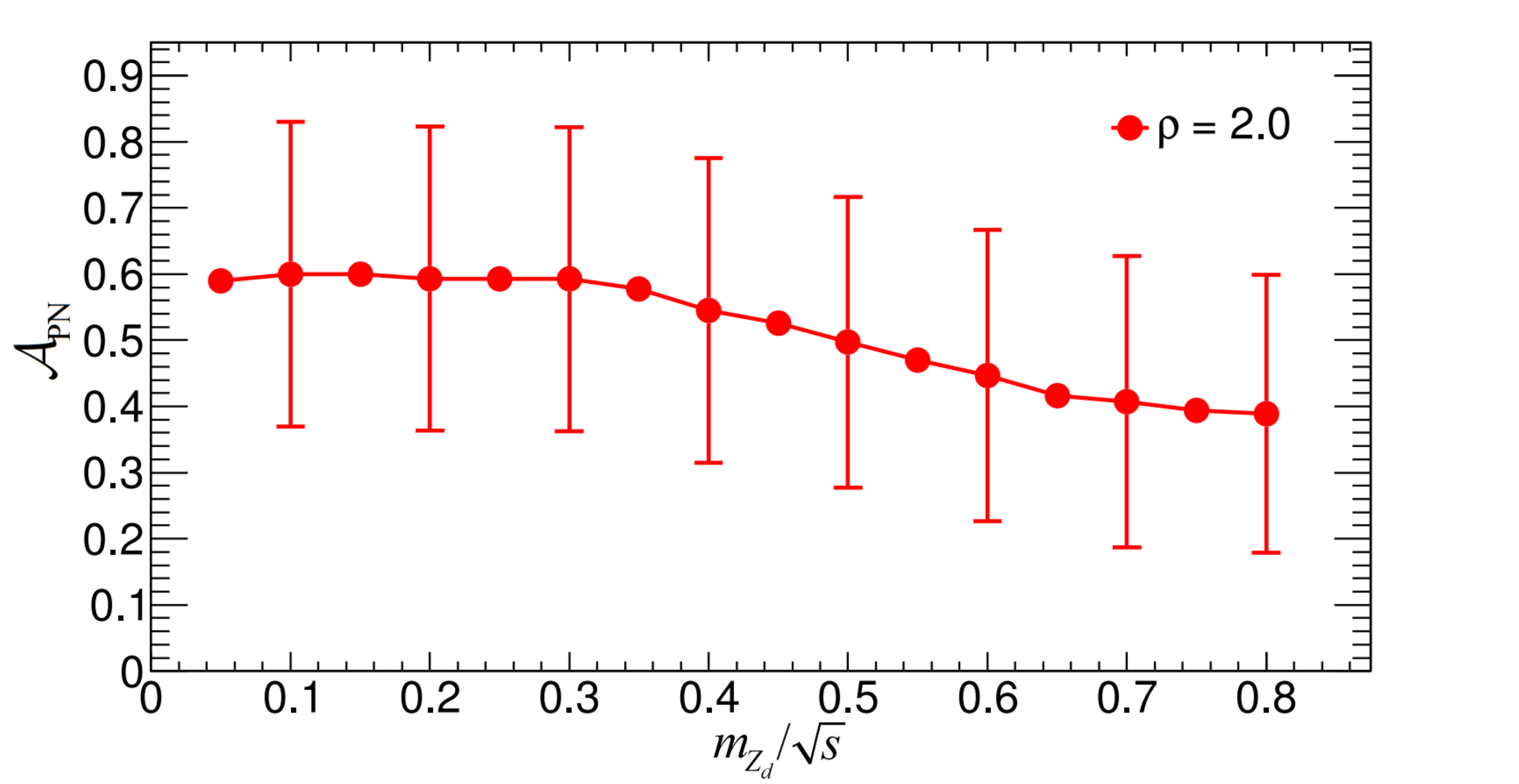}
		\caption{${\mathcal A}_{\rm PN}$ and its uncertainty as functions of $m_{Z_d}/\sqrt{s}$ for $S/\sqrt{B}=5$. We present results for $\rho=0.5$, $1.0$, $1.5$, and $2.0$, respectively.  }
		\label{APN_rho}
	\end{center}
\end{figure}
In Fig.~\ref{APN_rho}, we present results for $\rho=0.5$, $1.0$, $1.5$, and $2.0$, respectively. The results for other $\rho$'s can be easily inferred since ${\mathcal A}_{\rm PN}\propto \rho^2$.
It is seen that ${\mathcal A}_{\rm PN}$ is consistent with zero for $\rho=0.5$ and $\rho=1$, since the $1\sigma$ error bars in these cases reach ${\mathcal A}_{\rm PN}<0$ regime. On the other hand, for $\rho=2$, ${\mathcal A}_{\rm PN}$ are non-vanishing at more than $2\sigma$ for $m_{Z_d}/\sqrt{s}\leq 0.3$. 
 With $\mathcal{A}_{\rm PN}$ determined, the enhancement on detection significance 
 according to Eq.~(\ref{eq:sig_improve}) can be calculated. 
 Let us consider two special scenarios for $\rho$. The first scenario is setting $\varepsilon_\gamma=\varepsilon_Z$ in Eq.~(\ref{eq:INT}). Recasting the coupling of $Z_d$ to leptons into the standard form $e\varepsilon \bar{l}(g_{l,V}\gamma_{\mu}+g_{l,A}\gamma_{\mu}\gamma_5)lZ_d^{\mu}$, we have $\varepsilon=1.18\varepsilon_{\gamma}$, $g_{l,V}=-0.87$, and $g_{l,A}=-0.5$, which leads to $\rho=1.74$. 
 The second scenario is the extreme asymmetry case, such as $V-A$ interaction between $Z_d$ and SM fermions. This case is achieved with $\varepsilon_{\gamma}=\varepsilon_Z\tan\theta_W$ such that $\varepsilon=0.83\varepsilon_Z$ and $g_{l,A}=-g_{l,V}=1/\sqrt{2}$, which leads to $\rho=-2$. In this case the dark boson interacts with the up-type (down-type) quark the same way as it interacts with neutrino (charged lepton). 
 The values of $\mathcal{A}_{\rm PN}$ and the associated uncertainties in these two scenarios are summarized in 
 Table~\ref{result_summary} for $m_{Z_d}=0.5$ GeV and $2$ GeV. The improved detection significance with Eq.~(\ref{eq:sig_improve}) is also presented there. We note that, for $\vert \rho\vert=1.74$, the significance for a non-vanishing ${\mathcal A}_{\rm PN}$ already reaches $2\sigma$. 
  
It is of interest to estimate $S$ for a $5\sigma$ detection at $50$ ab$^{-1}$. Given the background cross sections $7.76\cdot 10^{-2}$ pb and
$2.54\cdot 10^{-2}$ pb for $m_{Z_d}=0.5 \ {\rm GeV}$ and $2 \ {\rm GeV}$, respectively, we have $B=3.88\cdot 10^6$ and $1.27\cdot 10^6$, respectively. For $S= 5\sqrt{B}$ assumed here, $S=9850$ and $5634$ for $m_{Z_d}=0.5 \ {\rm GeV}$ and $2 \ {\rm GeV}$, respectively. The way $S$ is split into $S_a$ and $S_b$ depends on $\mathcal{A}_{\rm PN}$. We present values of $S_a$ and $S_b$ for benchmark values of $m_{Z_d}$ and $\rho$ in Table~\ref{result_summary}. Finally, in the last two rows of Table~\ref{result_summary}, we also present ${\rm Br}(Z_d\to \mu^+\mu^-)$ and the corresponding $\varepsilon$ extracted   
from $S= 5\sqrt{B}$ at $50$ ab$^{-1}$.

\begin{table}[htbp]
	\begin{center}
		\begin{tabular}{lcccccc}\hline\hline  
			$\vert \rho \vert$    & \multicolumn{2}{c}{$0.00$} &  \multicolumn{2}{c}{$1.74$} &  \multicolumn{2}{c}{$2.00$} \\ \cline{2-7}
			$m_{Z_d}/{\rm GeV}$ & 0.5 & 2.0	& 0.5 & 2.0 & 0.5 & 2.0 \\ \hline
			${\mathcal A}_{\rm PN}$ &$(0.0\pm 0.2)$ & $(0.0\pm 0.2)$ & $(0.43\pm 0.22)$ & $(0.44\pm 0.22)$ & $(0.58\pm 0.23)$ & $(0.60\pm 0.23)$ \\ 
			Det. Sig. (Eq.~(\ref{eq:sig_improve})) & 5.0$\sigma$  & 5.0$\sigma$ & 5.4$\sigma$ & 5.5$\sigma$ & 5.8$\sigma$ & 5.8$\sigma$ \\
			$S(\kappa \cdot \xi>0)$ & 4925 & 2817 & 7040 & 4053 & 7780 & 4507 \\
			$S(\kappa \cdot \xi<0)$ & 4925 & 2817 & 2810 & 1581 & 2070 & 1127 \\	
			${\rm Br}(Z_d\to \mu^+\mu^-)$ & $40\%$ & $24\%$ & $21\%$ & $7.5\%$ & $17\%$ & $6.7\%$ \\ 
			$\varepsilon\cdot 10^4$  &3.3 & 3.2 & 4.6 & 5.7 & 5.1 & 6.1 \\
			\hline	
		\end{tabular}
	\end{center}
	\caption{Presented are the asymmetry parameter ${\mathcal A}_{\rm PN}$, the improved detection significance with two event bins, the signal event numbers, and 
	extracted $\varepsilon$ values assuming $5\sigma$ detection significance at $50$ ab$^{-1}$ based upon overall event excess, i.e., $S= 5\sqrt{B}$, for benchmark $m_{Z_d}$ and $\rho$ values. The $Z_d\to \mu^+\mu^-$ branching ratios are also presented.}	
\label{result_summary}
\end{table}
Before closing this section, we comment on $Z_d\to \mu^+\mu^-$ branching ratio for a non-vanishing $\rho$. It is clear that ${\rm Br}(Z_d\to \mu^+\mu^-)$ in the dark boson scenario differs from that of the dark photon case, in particular the decay $Z_d\to \nu\bar{\nu}$ is possible for the former case. From Eq.~(\ref{eq:INT}), one obtains $Z_d$ couplings to  leptons and quarks. For $\vert \rho\vert=1.74$ and $m_{Z_d}=0.5$ GeV, we have ${\rm Br}(Z_d\to e^+ e^-)=22\%$, ${\rm Br}(Z_d\to \mu^+\mu^-)=21\%$, $\sum_l {\rm Br}(Z_d\to \nu_l \bar{\nu}_l)=33\%$, and  ${\rm Br}(Z_d\to {\rm hadrons})=24\%$. For $m_{Z_d}=2$ GeV with the same $\vert \rho\vert$, the hadronic branching ratio increases to    
about $74\%$ due to the opening of $Z_d\to s\bar{s}$ channel. Furthermore, $\sum_l {\rm Br}(Z_d\to \nu_l \bar{\nu}_l)=11\%$ and ${\rm Br}(Z_d\to e^+ e^-)\simeq {\rm Br}(Z_d\to \mu^+\mu^-)=7.5\%$. 
For $\vert \rho\vert=2$, ${\rm Br}(Z_d\to \mu^+\mu^-)=17\%$ and $6.7\%$ for $m_{Z_d}=0.5$ GeV and $2$ GeV, respectively.  
We remark that our estimation of ${\rm Br}(Z_d\to {\rm hadrons})$ depends on the $Z_d$ mass. For $m_{Z_d}=2$ GeV, we use the quark level process $Z_d\to q\bar{q}$ to calculate the hadronic decay width. For $m_{Z_d}=0.5$ GeV, we consider only vector current contribution that can be inferred from the data as mentioned earlier. The axial-vector current is essentially not contributing since it cannot lead to two-pion final state due to parity conservation of strong interaction, nor its coupling to three-pion final state is noticeable at $0.5$ GeV from the study of axial-vector spectral function of hadronic $\tau$ decays~\cite{Davier:2005xq}.

\section{Summary and Conclusion}

In this article we have pointed out that the dark boson produced by $e^+e^-\to Z_d\gamma$ is transversely polarized in the limit $m_{Z_d}\ll \sqrt{s}$. This is a direct consequence of Ward-Takahashi identity. We also demonstrated this property by explicit calculations. The suppressed production of longitudinally-polarized dark boson state is shown in Fig.~\ref{fig:amp_square} for the $V-A$ limit, i.e., $g_{l,V}=-g_{l,A}=1/\sqrt{2}$. For $m_{Z_d}\ll \sqrt{s}$, the negative-helicity dark boson dominates the forward region ($\cos\theta>0$) while the positive-helicity one dominates the backward region ($\cos\theta <0$). As $m_{Z_d}$ approaches to $\sqrt{s}$, the production of longitudinally-polarized dark boson becomes noticeable. Furthermore, the angular distributions of negative- and positive-helicity dark bosons become indistinguishable.  

Since we aim for determining the parity violation parameter $\rho$, we analyze $\mu^-(\mu^+)$ angular distributions from polarized $Z_d$ decays. 
The double distribution of final state muons $d^2P/d\kappa d\xi$ ($\kappa=\cos\theta, \ \xi=\cos\theta_d$), defined in Eq.~(\ref{double_distribution}), was shown  to be sensitive to $\rho$. Explicitly we found that $d^2P/d\kappa d\xi=Q_0+Q_2\rho^2$ with $Q_0$ an even function of both $\kappa$ and $\xi$ and $Q_2$ an odd function of these variables. This implies that the signal event number in the kinematic range $\kappa\cdot \xi>0$ differs from that with $\kappa\cdot \xi< 0$, which motivates our definition of asymmetry parameter ${\mathcal A }_{\rm PN}$ proportional to $\rho^2$. Besides depending on $\rho^2$, ${\mathcal A }_{\rm PN}$ also depends  on the detector acceptance. We calculate numbers of signature and background events for two benchmark masses $m_{Z_d}=0.5$ GeV and $2$ GeV in Belle II detector. The resulting $90\%$ C.L. sensitivity to $\varepsilon$ at $500$ fb$^{-1}$ integrated luminosity is found to be consistent with that in Belle II physics book for the dark photon scenario. 

In the general scenario with non-vanishing $\rho$, we have seen that the detection significance of dark bosons increases by separately considering
events with different signs of $\kappa\cdot \xi$ rather than just counting the overall event excess. The increased $\chi^2$ value is proportional to ${\mathcal A }_{\rm PN}^2$, as seen from Eq.~(\ref{eq:sig_improve}). The numerical values of ${\mathcal A }_{\rm PN}$ are calculated as a function of $m_{Z_d}/\sqrt{s}$ for $\rho=0.5$, $1.0$, $1.5$, and $2.0$, respectively.  

In conclusion, we have shown that the detection of dark boson decays into muon pairs in $e^+ e^-$ colliders can probe the parity-violating couplings between the dark boson and SM fermions. 
Assuming a $5\sigma$ event excess in the search for $e^+e^-\to \gamma Z_d$ with $Z_d\to \mu^+\mu^-$ at Belle II, we have seen that the simultaneous fitting to event numbers in
positive and negative $\kappa\cdot \xi$ bins should improve the detection significance to $5.4\sigma$
and $5.8 \sigma$ for input true models with $\vert \rho\vert =1.74$ and  $\vert \rho\vert =2.0$, respectively. We have also seen that the significance for a non-vanishing ${\mathcal A}_{\rm PN}$ can reach $2\sigma$ for $\varepsilon_\gamma=\varepsilon_Z$ ($\rho=1.74$) with $m_{Z_d}/\sqrt{s}\ll 1$. The significance for general values of $m_{Z_d}/\sqrt{s}$ and $\rho$ can be inferred from Fig.~\ref{APN_rho} with suitable rescaling of the latter parameter. 

\begin{acknowledgments}
We thank P. Fayet for interesting comments. This work is supported by the Ministry of Science and
Technology, Taiwan under Grant No.~107-2119-M-009-017-MY3.
\end{acknowledgments}



\begin{thebibliography}{99}

\bibitem{Holdom:1985ag} 
  B.~Holdom,
  Phys.\ Lett.\  {\bf 166B}, 196 (1986).
  doi:10.1016/0370-2693(86)91377-8
 \bibitem{Galison:1983pa} 
  P.~Galison and A.~Manohar,
  Phys.\ Lett.\  {\bf 136B}, 279 (1984).
  doi:10.1016/0370-2693(84)91161-4
 \bibitem{Foot:2004pa} 
 R.~Foot,
 Int.\ J.\ Mod.\ Phys.\ D {\bf 13}, 2161 (2004)
 doi:10.1142/S0218271804006449
 [astro-ph/0407623].
 \bibitem{Feldman:2006wd} 
 D.~Feldman, B.~Kors and P.~Nath,
 Phys.\ Rev.\ D {\bf 75}, 023503 (2007)
 doi:10.1103/PhysRevD.75.023503
 [hep-ph/0610133].
 \bibitem{ArkaniHamed:2008qn} 
 N.~Arkani-Hamed, D.~P.~Finkbeiner, T.~R.~Slatyer and N.~Weiner,
 Phys.\ Rev.\ D {\bf 79}, 015014 (2009)
 doi:10.1103/PhysRevD.79.015014
 [arXiv:0810.0713 [hep-ph]].
 \bibitem{Pospelov:2008jd} 
 M.~Pospelov and A.~Ritz,
 Phys.\ Lett.\ B {\bf 671}, 391 (2009)
 doi:10.1016/j.physletb.2008.12.012
 [arXiv:0810.1502 [hep-ph]].
  \bibitem{Alexander:2016aln} 
  For a recent review, see J.~Alexander {\it et al.},
  arXiv:1608.08632 [hep-ph].
 \bibitem{Babu:1997st} 
 K.~S.~Babu, C.~F.~Kolda and J.~March-Russell,
 Phys.\ Rev.\ D {\bf 57}, 6788 (1998)
 doi:10.1103/PhysRevD.57.6788
 [hep-ph/9710441].
 \bibitem{Davoudiasl:2012ag} 
 H.~Davoudiasl, H.~S.~Lee and W.~J.~Marciano,
 Phys.\ Rev.\ D {\bf 85}, 115019 (2012)
 doi:10.1103/PhysRevD.85.115019
 [arXiv:1203.2947 [hep-ph]].
 \bibitem{Davoudiasl:2013aya} 
 H.~Davoudiasl, H.~S.~Lee, I.~Lewis and W.~J.~Marciano,
 Phys.\ Rev.\ D {\bf 88}, no. 1, 015022 (2013)
 doi:10.1103/PhysRevD.88.015022
 [arXiv:1304.4935 [hep-ph]].
 
\bibitem{Boehm:2003hm} 
  C.~Boehm and P.~Fayet,
  Nucl.\ Phys.\ B {\bf 683}, 219 (2004)
  doi:10.1016/j.nuclphysb.2004.01.015
  [hep-ph/0305261].

\bibitem{Borodatchenkova:2005ct} 
  N.~Borodatchenkova, D.~Choudhury and M.~Drees,
  Phys.\ Rev.\ Lett.\  {\bf 96}, 141802 (2006)
  doi:10.1103/PhysRevLett.96.141802
  [hep-ph/0510147].
 
\bibitem{Fayet:2007ua} 
  P.~Fayet,
  Phys.\ Rev.\ D {\bf 75}, 115017 (2007)
  doi:10.1103/PhysRevD.75.115017
  [hep-ph/0702176 [HEP-PH]].
 \bibitem{Batell:2009yf} 
 B.~Batell, M.~Pospelov and A.~Ritz,
 Phys.\ Rev.\ D {\bf 79}, 115008 (2009)
 doi:10.1103/PhysRevD.79.115008
 [arXiv:0903.0363 [hep-ph]].
 \bibitem{Essig:2009nc} 
 R.~Essig, P.~Schuster and N.~Toro,
 Phys.\ Rev.\ D {\bf 80}, 015003 (2009)
 doi:10.1103/PhysRevD.80.015003
 [arXiv:0903.3941 [hep-ph]].
\bibitem{Reece:2009un}
M.~Reece and L.~T.~Wang,
JHEP \textbf{07}, 051 (2009)
doi:10.1088/1126-6708/2009/07/051
[arXiv:0904.1743 [hep-ph]].
\bibitem{Essig:2013vha}
R.~Essig, J.~Mardon, M.~Papucci, T.~Volansky and Y.~M.~Zhong,
JHEP \textbf{11}, 167 (2013)
doi:10.1007/JHEP11(2013)167
[arXiv:1309.5084 [hep-ph]].
%
\bibitem{Karliner:2015tga}
M.~Karliner, M.~Low, J.~L.~Rosner and L.~T.~Wang,
Phys. Rev. D \textbf{92}, no.3, 035010 (2015)
doi:10.1103/PhysRevD.92.035010
[arXiv:1503.07209 [hep-ph]].
%
\bibitem{Araki:2017wyg} 
T.~Araki, S.~Hoshino, T.~Ota, J.~Sato and T.~Shimomura,
Phys.\ Rev.\ D {\bf 95}, no. 5, 055006 (2017)
doi:10.1103/PhysRevD.95.055006
[arXiv:1702.01497 [hep-ph]].

\bibitem{He:2017ord}
M.~He, X.~G.~He and C.~K.~Huang,
Int. J. Mod. Phys. A \textbf{32}, no.23n24, 1750138 (2017)
doi:10.1142/S0217751X1750138X
[arXiv:1701.08614 [hep-ph]].

\bibitem{He:2017zzr}
M.~He, X.~G.~He, C.~K.~Huang and G.~Li,
JHEP \textbf{03}, 139 (2018)
doi:10.1007/JHEP03(2018)139
[arXiv:1712.09095 [hep-ph]].

\bibitem{Jiang:2018jqp}
J.~Jiang, H.~Yang and C.~F.~Qiao,
Eur. Phys. J. C \textbf{79} (2019) no.5, 404
doi:10.1140/epjc/s10052-019-6912-3
[arXiv:1810.05790 [hep-ph]].
%
\bibitem{Babusci:2014sta} 
  D.~Babusci {\it et al.} [KLOE-2 Collaboration],
  Phys.\ Lett.\ B {\bf 736}, 459 (2014)
  doi:10.1016/j.physletb.2014.08.005
  [arXiv:1404.7772 [hep-ex]].

\bibitem{Lees:2014xha} 
  J.~P.~Lees {\it et al.} [BaBar Collaboration],
  Phys.\ Rev.\ Lett.\  {\bf 113}, no. 20, 201801 (2014)
  doi:10.1103/PhysRevLett.113.201801
  [arXiv:1406.2980 [hep-ex]].
  
\bibitem{Anastasi:2015qla} 
  A.~Anastasi {\it et al.},
  Phys.\ Lett.\ B {\bf 750}, 633 (2015)
  doi:10.1016/j.physletb.2015.10.003
  [arXiv:1509.00740 [hep-ex]].
\bibitem{Anastasi:2016ktq} 
  A.~Anastasi {\it et al.} [KLOE-2 Collaboration],
  Phys.\ Lett.\ B {\bf 757}, 356 (2016)
  doi:10.1016/j.physletb.2016.04.019
  [arXiv:1603.06086 [hep-ex]].
 
\bibitem{Ablikim:2017aab} 
  M.~Ablikim {\it et al.} [BESIII Collaboration],
  Phys.\ Lett.\ B {\bf 774}, 252 (2017)
  doi:10.1016/j.physletb.2017.09.067
  [arXiv:1705.04265 [hep-ex]].

\bibitem{Anastasi:2018azp} 
  A.~Anastasi {\it et al.} [KLOE-2 Collaboration],
  Phys.\ Lett.\ B {\bf 784}, 336 (2018)
  doi:10.1016/j.physletb.2018.08.012
  [arXiv:1807.02691 [hep-ex]].
  
\bibitem{Lees:2017lec} 
  J.~P.~Lees {\it et al.} [BaBar Collaboration],
  Phys.\ Rev.\ Lett.\  {\bf 119}, no. 13, 131804 (2017)
  doi:10.1103/PhysRevLett.119.131804
  [arXiv:1702.03327 [hep-ex]].
  
\bibitem{WT}
J.~C.~ Ward, Phys. \ Rev. {\bf 78}, 1824 (1950); Y.~Takahashi, Nuovo \ Cimento {\bf 6}, 370 (1957).
\bibitem{Kou:2018nap}
E.~Kou \textit{et al.} [Belle-II],
PTEP \textbf{2019} (2019) no.12, 123C01
doi:10.1093/ptep/ptz106
[arXiv:1808.10567 [hep-ex]].

\bibitem{Abe:2010gxa} 
  T.~Abe {\it et al.} [Belle-II Collaboration],
  arXiv:1011.0352 [physics.ins-det].

\bibitem{Brodzicka:2012jm} 
  J.~Brodzicka {\it et al.} [Belle Collaboration],
  PTEP {\bf 2012}, 04D001 (2012)
  doi:10.1093/ptep/pts072
  [arXiv:1212.5342 [hep-ex]].
 
\bibitem{Bevan:2014iga} 
  A.~J.~Bevan {\it et al.} [BaBar and Belle Collaborations],
  Eur.\ Phys.\ J.\ C {\bf 74}, 3026 (2014)
  doi:10.1140/epjc/s10052-014-3026-9
  [arXiv:1406.6311 [hep-ex]].
  
\bibitem{Feng:2016jff}
J.~L.~Feng, B.~Fornal, I.~Galon, S.~Gardner, J.~Smolinsky, T.~M.~P.~Tait and P.~Tanedo,
 Phys.\ Rev.\ Lett.\  {\bf 117}, no. 7, 071803 (2016)
 doi:10.1103/PhysRevLett.117.071803
 [arXiv:1604.07411 [hep-ph]].

\bibitem{Feng:2016ysn} 
J.~L.~Feng, B.~Fornal, I.~Galon, S.~Gardner, J.~Smolinsky, T.~M.~P.~Tait and P.~Tanedo,
Phys.\ Rev.\ D {\bf 95}, no. 3, 035017 (2017)
doi:10.1103/PhysRevD.95.035017
[arXiv:1608.03591 [hep-ph]].

\bibitem{Feng:2020mbt}
J.~L.~Feng, T.~Tait, M.P. and C.~B.~Verhaaren,
[arXiv:2006.01151 [hep-ph]].

\bibitem{Krasznahorkay:2015iga} 
 A.~J.~Krasznahorkay {\it et al.},
 Phys.\ Rev.\ Lett.\  {\bf 116}, no. 4, 042501 (2016)
 doi:10.1103/PhysRevLett.116.042501
 [arXiv:1504.01527 [nucl-ex]].

\bibitem{Krasznahorkay:2019lyl}
A.~J.~Krasznahorkay, M.~Csatlós, L.~Csige, J.~Gulyás, M.~Koszta, B.~Szihalmi, J.~Timár, D.~S.~Firak, Á.~Nagy, N.~J.~Sas and G.~Cern,
[arXiv:1910.10459 [nucl-ex]].

\bibitem{Jiang:2018uhs} 
 J.~Jiang, L.~B.~Chen, Y.~Liang and C.~F.~Qiao,
  Eur.\ Phys.\ J.\ C {\bf 78}, no. 6, 456 (2018).
  doi:10.1140/epjc/s10052-018-5945-3

\bibitem{Alikyhanov:2017cp} 
I.~Alikhanov and E.~A.~Paschos,
Phys.\ Rev.\ D {\bf 97}, no. 11, 115004 (2018)
doi:10.1103/PhysRevD.97.115004
[arXiv:1710.10131 [hep-ph]].
 \bibitem{Fayet:1977yc}
 P.~Fayet,
 Phys. Lett. B \textbf{69}, 489 (1977)
 doi:10.1016/0370-2693(77)90852-8
 \bibitem{Fayet:1980ss}
 P.~Fayet,
 Phys. Lett. B \textbf{96}, 83-88 (1980)
 doi:10.1016/0370-2693(80)90217-8
 \bibitem{Bouchiat:1974kt}
 M.~A.~Bouchiat and C.~C.~Bouchiat,
 Phys. Lett. B \textbf{48}, 111-114 (1974)
 doi:10.1016/0370-2693(74)90656-X
 \bibitem{Bouchiat:1979cq}
 C.~Bouchiat, Proc. Workshop on Neutral Current Interactions in Atoms (Carg\`ese, France, September 1979)(Univ. of Michigan Press).
 %
\bibitem{Abdullah:2018ykz}
M.~Abdullah, J.~B.~Dent, B.~Dutta, G.~L.~Kane, S.~Liao and L.~E.~Strigari,
Phys. Rev. D \textbf{98}, no.1, 015005 (2018)
doi:10.1103/PhysRevD.98.015005
[arXiv:1803.01224 [hep-ph]].
%
\bibitem{Tanabashi:2018oca}
M.~Tanabashi \textit{et al.} [Particle Data Group],
Phys. Rev. D \textbf{98}, no.3, 030001 (2018)
doi:10.1103/PhysRevD.98.030001
\bibitem{Adachi:2018qme}
I.~Adachi \textit{et al.} [Belle-II],
Nucl. Instrum. Meth. A \textbf{907}, 46-59 (2018)
doi:10.1016/j.nima.2018.03.068
\bibitem{Belyaev:2012qa}
A.~Belyaev, N.~D.~Christensen and A.~Pukhov,
Comput. Phys. Commun. \textbf{184}, 1729-1769 (2013)
doi:10.1016/j.cpc.2013.01.014
[arXiv:1207.6082 [hep-ph]].
%
\bibitem{Davier:2005xq}
For a comprehensive review, see 
M.~Davier, A.~Hocker and Z.~Zhang,
Rev. Mod. Phys. \textbf{78}, 1043-1109 (2006)
doi:10.1103/RevModPhys.78.1043
[arXiv:hep-ph/0507078 [hep-ph]].




  
  

  



 










\end{thebibliography}
\end{document}